\documentclass[aps,10pt,prl,twocolumn,amsmath,amssymb,superscriptaddress]{revtex4-1}

\usepackage[english]{babel} 
\usepackage[utf8]{inputenc}
\usepackage[T1]{fontenc}

\usepackage{bbm}
\usepackage{verbatim}
\usepackage{graphicx}
\usepackage{times}
\usepackage{epsfig}
\usepackage{graphicx}
\usepackage{bm}
\usepackage{txfonts}
\usepackage{dsfont}
\usepackage{color}

\usepackage{hyperref}
\hypersetup{
colorlinks=true,
linkcolor=blue,          
citecolor=blue,          
filecolor=magenta,       
urlcolor=cyan
}

\newcommand{\ket}[1]{|#1\rangle}
\newcommand{\bra}[1]{\langle#1|}

\newcommand{\fa}{\alpha}

\newcommand{\fpa}{\partial_\alpha}
\newcommand{\fpas}{{\partial_{\alpha^*}}}

\begin{document}

\selectlanguage{english}

\title{Genuine quantum signatures in synchronization of anharmonic self-oscillators}

\author{Niels L\"orch}
\affiliation{Department of Physics, University of Basel, Klingelbergstrasse 82, CH-4056 Basel, Switzerland}

\author{Ehud Amitai}
\affiliation{Department of Physics, University of Basel, Klingelbergstrasse 82, CH-4056 Basel, Switzerland}

\author{Andreas Nunnenkamp}
\affiliation{Cavendish Laboratory, University of Cambridge, Cambridge CB3 0HE, United Kingdom}

\author{Christoph Bruder}
\affiliation{Department of Physics, University of Basel, Klingelbergstrasse 82, CH-4056 Basel, Switzerland}

\date{\today}

\begin{abstract}
We study the synchronization of a van der Pol self-oscillator with Kerr anharmonicity to an external drive. We demonstrate that the anharmonic, discrete energy spectrum of the quantum oscillator leads to multiple resonances in both phase locking and frequency entrainment not present in the corresponding classical system. Strong driving close to these resonances leads to nonclassical steady-state Wigner distributions. Experimental realizations of these genuine quantum signatures can be implemented with current technology.
\end{abstract}

\pacs{}

\maketitle

Synchronization of self-oscillators is a subject with great relevance to several natural sciences \cite{Balanov2008, Pikovsky2015}. Its exciting frontiers include neuronal synchronization in the human brain \cite{Varela2001, Fell2011}, stabilization of power-grid networks \cite{Motter2013}, as well as the engineering of high-precision clocks \cite{Antonio2012, Arroyo2013}. Recent advances in nanotechnology will enable experiments with large arrays of self-oscillators in the near future~\cite{Dykman2012, Safavi-Naeini2014a}. Whereas most research has focused on the classical domain, synchronization in the quantum regime \cite{Shepelyansky2006} has become a very active topic. There has been much recent experimental progress with micro- and nanomechanical systems \cite{Zhang2012, Bagheri2013, Matheny2014, Shlomi2015, Zhang2015}, and theoretical proposals for mesoscopic ensembles of atoms \cite{Xu2014b, Xu2015, Xu2015c}, lasers \cite{Carmele2013}, cavity optomechanics \cite{Ludwig2013, Walter2014a, Walter2014, Weiss2015}, trapped ions \cite{Lee2013, Lee2014a, Hush2015}, arrays of coupled nonlinear cavities \cite{Jin2013}, and interacting quantum dipoles \cite{Zhu2015}. In addition, there are open conceptual questions on the relation of synchronization to entanglement or mutual information \cite{Fazio2013, Fazio2015}.

Studying a van der Pol oscillator, the most prominent example of a self-oscillator, recent theoretical work characterized how synchronization quantitatively differs between its quantum and classical realization in phase locking \cite{Lee2013, Lee2014a} as well as in frequency entrainment \cite{Walter2014a, Walter2014}. While synchronization is hindered by quantum noise compared to the classical model \cite{Walter2014a, Walter2014}, noise is less detrimental \cite{Lee2013, Lee2014a} than one would expect from a semiclassical description.

In this Letter we study self-oscillators for which \textit{both} the damping and the frequency is amplitude-dependent. We show that their synchronization behavior is \emph{qualitatively} different in the quantum and the classical regime. Focusing on a van der Pol oscillator with Kerr anharmonicity, we find two genuine quantum signatures. First, while synchronization of one such oscillator to an external drive is maximal at one particular frequency classically, the corresponding quantum system shows a tendency to synchronize at multiple frequencies. Using perturbation theory in the drive strength, we demonstrate that these multiple resonances reflect the quantized anharmonic energy spectrum of the oscillator. We show that these features are observable in the phase probability distribution if the Kerr anharmonicity is large compared to the relaxation rates and the system is in the quantum regime, i.e.~the limit cycle amplitudes are small. In the semiclassical limit the energy spectrum becomes continuous, so that the resonances (and therefore the quantized energy spectrum) cannot be resolved. Using numerically exact simulations of the full quantum master equation, we find a second genuine quantum signature: for strong driving close to these resonances the steady-state Wigner distribution exhibits areas of negative density, i.e., the steady state is nonclassical.

\begin{figure}[tb]
\centering
\includegraphics[width=0.5\textwidth]{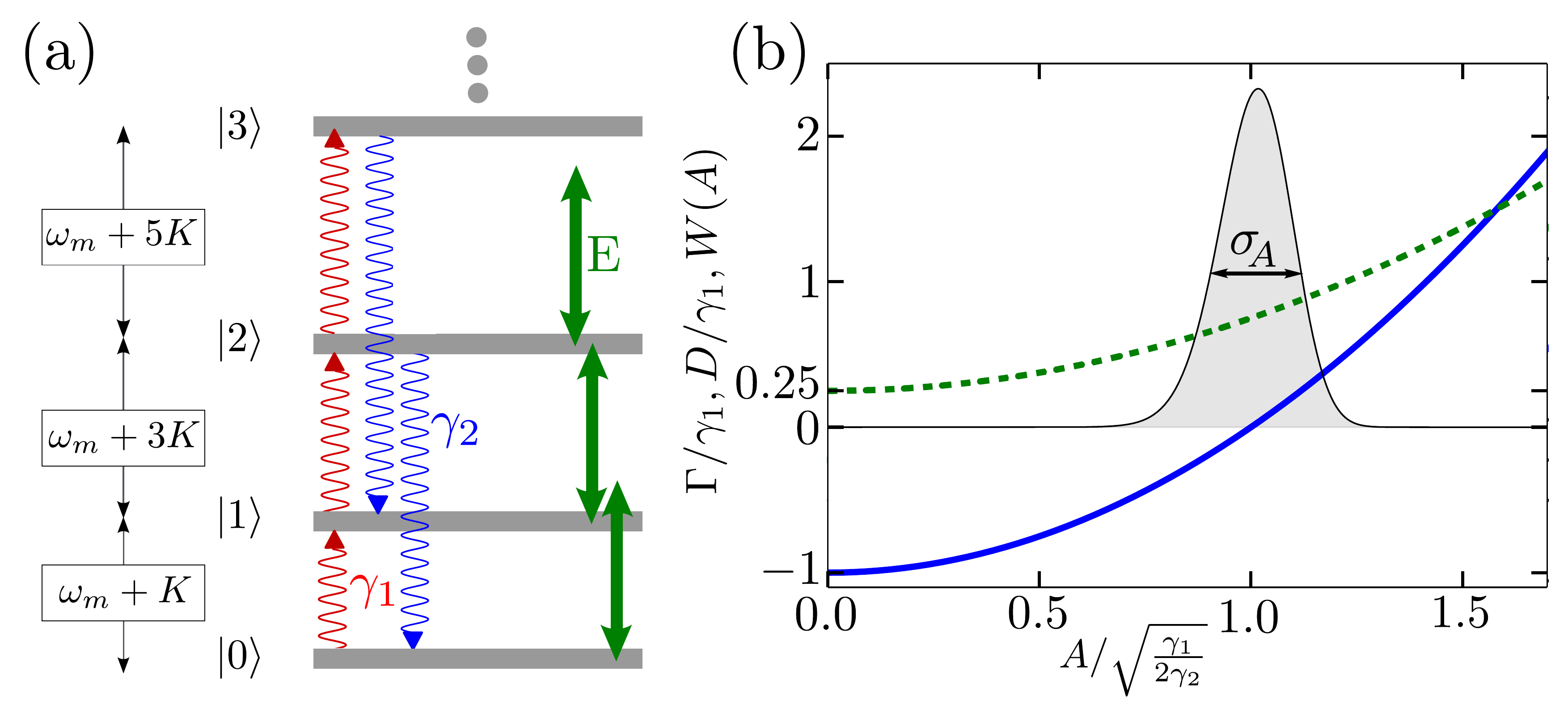}
\caption{
Van der Pol self-oscillator with Kerr anharmonicity. (a) Lowest-lying energy levels. The Kerr anharmonicity $K$ leads to a level spacing $\omega_m+(2n+1)K$ increasing with excitation number $n$. In this figure the coherent drive (green arrows) is resonant with the transition between the states $\ket 1$ and $\ket 2$. The wiggly lines denote two incoherent processes: linear (one-phonon) anti-damping with rate $\gamma_1$ (red arrows) and nonlinear (two-phonon) damping with rate $\gamma_2$ (blue arrows). (b) The amplitude-dependent damping rate $\Gamma$ (blue solid line) and amplitude-dependent diffusion constant $D$ (green dashed line) in the semiclassical equation \eqref{determ} as a function of the amplitude $A$. In the limit of large amplitude $A$, the radial Wigner density $W(A)$ is a Gaussian with variance $\sigma_A^2=3/8$ around the zero of $\Gamma$, i.e.~$\Gamma(A) = 0$.}
\label{AnalytFigurePrel}
\end{figure}

\textit{Model}.--
We consider an anharmonic self-oscillator subject to an external drive.
For concreteness, we will focus on a van der Pol self-oscillator with Kerr anharmonicity, 
but the results we present are generic and can be generalized to other anharmonic self-oscillators. 
In the rotating frame of the drive our model system is described by the quantum master equation
\begin{align}
&\dot \rho =  -i[H_0+H_1, \rho] + L \rho, \label{meq} 
\end{align}
with Hamiltonian $H_0= -\Delta a^\dagger a  + K (a^\dagger a)^2$, drive Hamiltonian $H_1 = i E (a - a^\dagger)$, and Lindblad operator $L \rho=\frac {\gamma_1} 2 \mathcal D[a^\dagger] \rho + \frac {\gamma_2} 2 \mathcal D[a^2] \rho$, where $a$ denotes the annihilation operator for the oscillator and $\mathcal D[x]  \rho=2x \rho x^\dagger - (x^\dagger x \rho -\rho x^\dagger x)$. The Hamiltonian $H_0$ describes a Kerr oscillator with anharmonic spectrum characterized by the Kerr parameter $K>0$, see Fig.~\ref{AnalytFigurePrel} (a). The coherent drive has amplitude $E$ and frequency $\omega_d$ that is detuned from the (harmonic) frequency of the oscillator $\omega_m$ by $\Delta = \omega_d-\omega_m$. The oscillator is also subject to two incoherent processes described by the Lindblad operator $L$, i.e.~linear (one-phonon) anti-damping with rate $\gamma_1$ and nonlinear (two-photon) damping with rate $\gamma_2$.

\textit{Phase space description and semiclassical model.--}
Fully equivalent to the quantum master equation~\eqref{meq}, the system can be described by a partial differential equation $\partial_t W(\alpha,\alpha^*,t) = \Lambda W(\alpha,\alpha^*,t)$ for the Wigner distribution $W$ \cite{Gardiner2004b,Carmichael} with
\begin{align}
\Lambda W &=  \frac{\gamma_1}2 ( -\fpa \fa  + \tfrac 12 \fpas \fpa )  W \nonumber\\ 
&+ \frac {\gamma_2}2 \left[ 2 \fpa \fa (|\alpha|^2-1) +  \fpas \fpa (2|\alpha|^2-1) + \tfrac 12  \fpa^2 \fpas \alpha  \right] W \nonumber \\
&+ i K \left[ \fpa \fa (2 |\alpha|^2 -1) -\tfrac 14 \fpa^2 \fpas \fa \right] W \nonumber \\
&- i \Delta  \fpa \fa  W
+ E \fpa W 
+\mathrm{h.c.}
\label{qPDG}
\end{align}
Both the Kerr anharmonicity $K$ and the van der Pol nonlinearity $\gamma_2$ lead to third-order derivatives in $\alpha$ that are necessary for nonclassical steady-state Wigner densities \cite{Risken1984}. Indeed, for the van der Pol term $\gamma_2$ the third-order derivative is accompanied with a diffusion term limiting the genuine quantum behavior, whereas the Kerr term $K$ gives us the opportunity to increase `quantumness' without adding diffusion. 
Note that this could be equally well achieved with other anharmonic Hamiltonian terms, stemming e.g. from an anharmonic Duffing potential \cite{Katz2008, Supplemental}.

In the limit of large limit-cycle amplitudes $|\alpha|$, i.e.~$\gamma_1 \gg \gamma_2$, we can neglect the third-order derivatives \cite{DeFaria2005, Lee2013, Walls2007} and get
\begin{align}
\Lambda_c = \fpa \left[\left( \frac {\Gamma (|\alpha|)}2 +i  \Omega (|\alpha|) \right) \alpha +E \right] + \fpas \fpa D(|\alpha|) +\mathrm{h.c.},
\label{determ}
\end{align}
that contains only first- and second-order derivatives corresponding to drift and diffusion, respectively. As illustrated in Fig.~\ref{AnalytFigurePrel} (b), the drift term consists of an amplitude-dependent damping rate $\Gamma=-\gamma_1+2\gamma_2 (|\alpha|^2-1)$, an amplitude-dependent oscillation frequency $\Omega=-\Delta+2K(|\alpha|^2-1)$ in the frame of the drive, and the drive of strength $E$. The diffusion is given by $D= \frac{\gamma_1}4+\frac {\gamma_2} 2(2|\alpha|^2-1)$.

\begin{figure}
\centering
\includegraphics[width=0.5\textwidth]{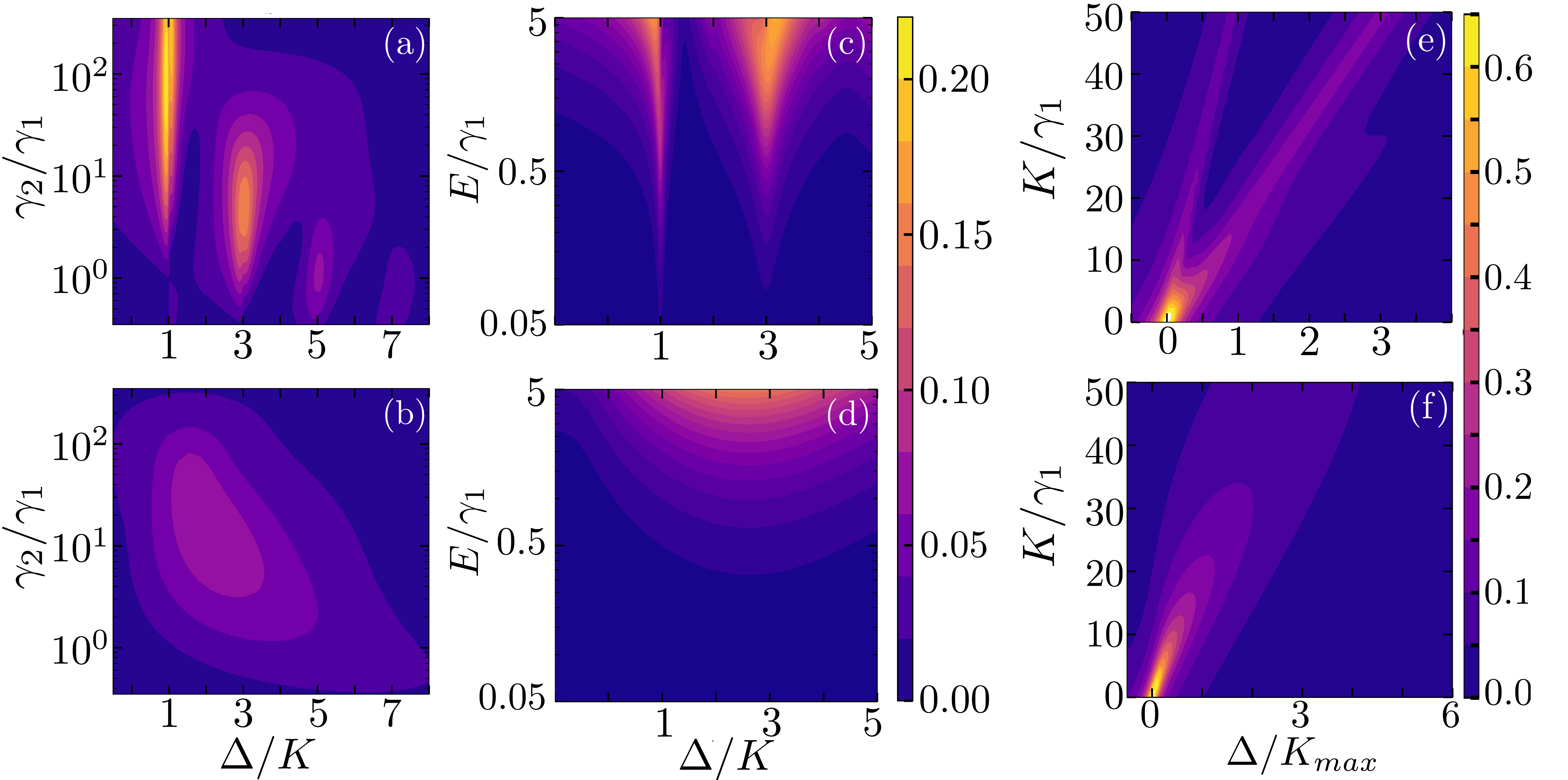}
\caption{Global behavior of the phase-locking measure $|S|$ for the steady state of $\Lambda$ (a, c, e) and $\Lambda_c$ (b, d, f).  In (a) and (b) $|S|$ is plotted as a function of $\Delta$ and $\gamma_2$ for $E=2.25 \gamma_1$ and $K=50 \gamma_1$, in (c) and (d) as a function of $\Delta$ and $E$ for $\gamma_2=5 \gamma_1$ and $K=50 \gamma_1$, and in (e) and (f) as a function of $\Delta$ and $K$ for $\gamma_2=5 \gamma_1$, $E=2.25 \gamma_1$, and $K_\textrm{max}= 50 \gamma_1$.}
\label{Overview}
\end{figure}

In the absence of driving $E=0$ and using polar coordinates $\alpha = A e^{i\phi}$, the dynamics of the amplitude $A$ decouples from the dynamics of the phase $\phi$ in Eq.~(\ref{determ}). Within a Gaussian approximation similar to Refs.~\cite{Rodrigues2010, Lorch2014} we solve for the radial steady-state distribution $W(A)$ and find a mean amplitude $A_0=\sqrt{1+{\gamma_1}/{2\gamma_2}}$ for which $\Gamma(A_0)=0$. For $A\gg 1$ we obtain a variance $\sigma^2_A=\frac 3 8$ so that the relative deviation $\sigma_A /A_0$ is negligible and we can approximate the amplitude-dependent diffusion constant with its value at $A_0$, i.e.~$D \approx ( 3\gamma_1 +2\gamma_2)/4>0$. In this case, $\Lambda_c$ is a Fokker-Planck-operator describing a classical process. The oscillation frequency $\Omega$ is sensitive to fluctuations in the amplitude $A$, i.e.~$\sigma_\Omega \propto K A_0 \approx K \sqrt{\gamma_1/2\gamma_2}$. Therefore, classically, the range of detuning $\Delta$ for which phase locking and frequency entrainment occur becomes larger with increasing $K$ and $A_0$, as we shall also see in Figs.~\ref{Overview} (b) and (f).

\textit{Analytical treatment}.-- To gain some analytical understanding, we use perturbation theory \cite{Li2014} to approximate the steady state of the quantum master equation \eqref{meq} in the limit of weak drive strength and large Kerr anharmonicity $E \ll \gamma_1+\gamma_2 \ll K$. In analogy to standard perturbation theory for Hamiltonians, we decompose the quantum master equation $\dot \rho= \left( \mathcal L_0 + \mathcal L_1 \right) \rho$ in Eq.~\eqref{meq} into an unperturbed operator $\mathcal L_0$ and a perturbation $\mathcal L_1$ with $\mathcal L_0 \, \rho = L \rho -i [H_0, \rho] $ and $\mathcal L_1 \, \rho = -i [H_1, \rho] $. The first-order correction to the steady state is $\rho^{(1)}=-\mathcal L_0^{-1} \mathcal L_1 \rho^{(0)}$ where $\rho^{(0)}$ is the steady state of the unperturbed Liouvillian $\mathcal L_0$ and $\mathcal L_0^{-1}$ is its Moore-Penrose pseudoinverse.

The unperturbed steady-state $\rho^{(0)}$ can be found analytically: $\rho^{(0)}_{nn}=r^n \Phi(1+n,r+n,r)/[(r)_n \Phi(1,r,2r)]$ where $(\cdot)_n$ denotes the Pochhammer symbol, $\Phi$ is Kummer's confluent hypergeometric function, and $r=\gamma_1/\gamma_2$ \cite{Dodonov1997}. We see that $\rho^{(0)}$ is diagonal in the number basis describing limit cycles without any preferred phase, i.e.~their Wigner density is rotationally symmetric, and it depends only on the ratio of relaxation rates $\gamma_1/\gamma_2$ and not the Kerr parameter $K$. In the limit of large $r$, corresponding to large mean amplitude,  the $ \rho^{(0)}_{nn}$ follow a Gaussian distribution with mean $\langle n \rangle = r/2$ and variance $\Delta^2 n =3r/4$. This is consistent with the large-amplitude semiclassical treatment above, as both mean $\langle n \rangle \approx A_0^2$ and Fano factor $\Delta^2 n/\langle n \rangle \approx 4 \sigma_A^2$ agree. In the opposite limit $r \to 0$, the steady state is approximately $ \rho^{(0)} \to \frac 23 \ket 0 \bra 0 + \frac 13 \ket 1 \bra 1 + \mathcal O (\gamma_1/\gamma_2)$.

Next, we exploit the fact that the superoperator $\mathcal L_0$ can be decomposed into a term coupling diagonal density matrix elements and a term coupling off-diagonal elements separately. Neglecting terms of order $\gamma_1/K$ and $\gamma_2/K$, we obtain the inverse $\mathcal L_0^{-1}$ in the off-diagonal subspace by inverting its diagonal so that $\mathcal L_0^{-1} \ket{m+1} \bra m \approx \lambda_{m+1,m}^{-1}\ket{m+1} \bra m$ with 
\begin{align}
\lambda_{m+1,m}=  i \left[ \Delta -K (2m+1) \right] -\frac {\Gamma_m}2\:,
\end{align}
where
\begin{align}
\Gamma_m=\gamma_1 (2m+3) + 2 \gamma_2 m^2\:. 
\label{GammaM}
\end{align}

Finally, as $\mathcal L_1$ couples only neighboring Fock states, $\rho^{(1)}=-\mathcal L_0^{-1} \mathcal L_1 \rho^{(0)}$ has nonzero elements only on the minor diagonals, so that the first-order correction for the steady state is
\begin{align}
\rho^{(1)}_{m+1,m} = \rho^{(1)\:*}_{m,m+1} =
\sqrt{m+1}E \frac{\left( \rho^{(0)}_{mm}-\rho^{(0)}_{m+1,m+1}\right)}{\lambda_{m+1,m}}\:. 
\label{rho_s}
\end{align}
The anharmonic quantum energy levels of the Kerr oscillator shown in Fig.~\ref{AnalytFigurePrel} (a) lead to multiple resonances in the first-order response to an external drive. In the following we will discuss the consequences of these resonances in terms of phase locking and frequency entrainment.

\textit{Phase locking}.-- 
It is well-known that there is not a unique way to define the phase operator in quantum mechanics \cite{gerry2005}. One option \cite{Barak2005} that has been used to study quantum synchronization \cite{Hush2015} is the phase distribution $\mathcal P(\phi) = \tfrac{1}{2\pi}\bra \phi \rho \ket \phi$ with $\ket \phi=\sum_{n=0}^\infty e^{in \phi} \ket n$, yielding $2\pi\mathcal P(\phi)-1=\sum_{m\ne n=0}^\infty \rho_{m,n} e^{i\phi(n-m)}$. 
Our perturbative steady-state solution \eqref{rho_s} contains only terms with $n-m=\pm 1$, so $\mathcal P(\phi) = \tfrac{1}{2\pi} + \eta_1 \cos \phi + \eta_2 \sin \phi$ with $\eta_1 = \frac{1}{\pi}\sum_{m=0}^\infty \textrm{Re}[\rho_{m+1,m}]$ and $\eta_2 = \frac{1}{\pi}\sum_{m=0}^\infty \textrm{Im}[\rho_{m+1,m}]$.

To convert the phase distribution $\mathcal P(\phi)$ into a single number characterizing the tendency to synchronize, we use the absolute value of the measure defined in Ref.~\cite{Weiss2015}, i.e.
\begin{align}
S=|S|e^{i\theta}=\frac{\langle a \rangle} { \sqrt{\langle a^\dagger a \rangle}}=\frac{\sum_{m=0}^\infty \sqrt{m+1} \rho_{m+1,m} }{\sqrt{ \sum_{m=0}^\infty m \rho_{m,m}} }\:.
\label{measure1}
\end{align}
{ Note that the Hamiltonian in \eqref{meq} is
  time-independent, as it is written already in the rotating frame of
  the external drive. As a
  consequence, $S$ is also independent of time in steady state.  The
  mean relative phase between the external drive and the
  self-oscillator is measured by $\theta$.  For synchronization to
  occur it suffices that there is only a small {\it variation} of
  the phase.  Such a small variation leads to large values of $|S|
  \to 1$, which we therefore adopt as our synchronization measure
  to compare phase locking in the quantum case (\ref{qPDG}) and the
  semiclassical case (\ref{determ}).
  The exact value of the relative phase $\theta$ is not relevant
  for our purposes and is therefore discarded here.}

\begin{figure}
\centering
\includegraphics[width=0.5\textwidth]{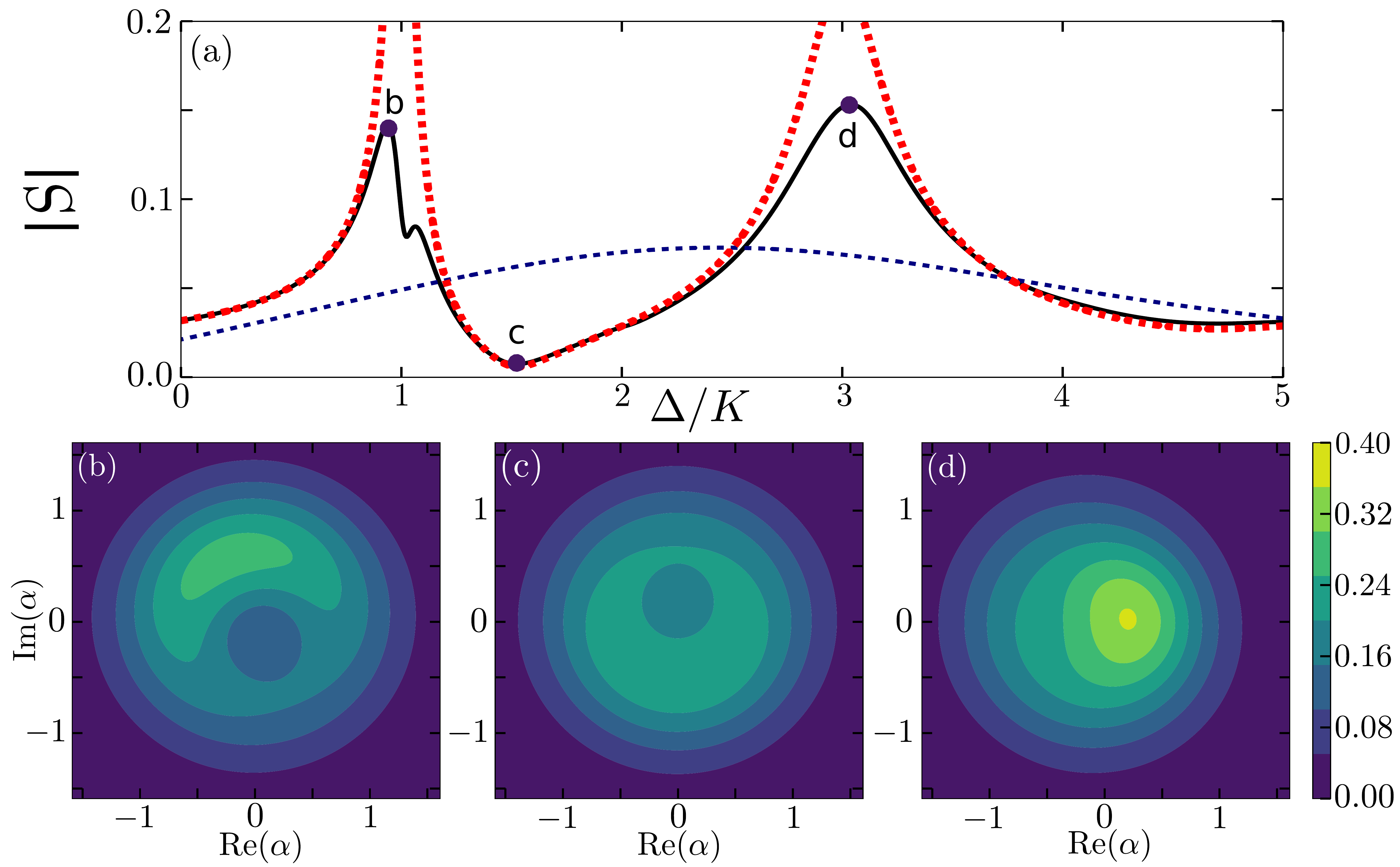}
\caption{(a) Phase locking measure $|S|$ for forced synchronization and corresponding Wigner distributions. The phase locking behavior for the quantum system (black solid line) described by $\Lambda$ (defined in Eq.~\eqref{qPDG}) can be understood with our perturbative expression (\ref{Lorentzref}) (red bold dotted line). For the parameters of this plot ($\gamma_2/\gamma_1=7$, $E/\gamma_1=2.25$, $K/\gamma_1=50$) approximately three energy levels have significant occupation, so that two resonances are possible. The blue dashed line shows the results of the corresponding semi-classical model $\Lambda_c$ (defined in Eq.~\eqref{determ}), for which there is only one resonance as expected. The {time-independent steady-state} Wigner distributions for the parameters at the two peaks (b), (d) and the minimum (c) illustrate the quantum phase locking behavior of $\Lambda$.}
\label{AnalytFigure}
\end{figure}

Evaluating $S$ for the perturbative steady-state solution (\ref{rho_s}), we obtain
\begin{align}
S(\rho^{(1)}) =   \sum_{m=0}^{\infty}  \left( \rho^{(0)}_{m+1,m+1} - \rho^{(0)}_{mm} \right) \frac{m+1} {\sqrt{\langle a^\dagger a \rangle}} \frac E {\lambda_{m+1,m}} \:.
\label{Lorentzref}
\end{align}
Equation~\eqref{Lorentzref} is one of the main results of this Letter. $S(\rho^{(1)})$ is a coherent sum of resonances at $\Delta=K(2m+1)$ and width $\Gamma_m$. They can be resolved for large Kerr anharmonicity $K \gg \Gamma_m\:$ defined in Eq.~(\ref{GammaM}). The number of visible resonances depends on the number of non-negligible probabilities $\rho^{(0)}_{mm}$ in the unperturbed steady state $\rho^{(0)}$. In the quantum limit $r\rightarrow 0$,  the resonances become more pronounced since fewer levels are occupied. In the limit $r\rightarrow \infty$, the energy spectrum becomes continuous, so that the resonances can no longer be resolved.

With this analytical understanding in mind we now present exact numerical steady-state results of Eq.~(\ref{meq}) obtained with the steady-state solver of QuTiP \cite{Johansson2011b, Johansson2013} which we compare to the semiclassical approximation described by $\Lambda_c$ of Eq.~\eqref{determ}, where the steady-state is found by discretizing the Fokker-Planck equation. In Figs.~\ref{AnalytFigure} and \ref{Overview} the resulting phase-locking measure $|S|$ is plotted as a function of the system parameters. The black solid line in Fig.~\ref{AnalytFigure} (a) shows $|S|$ as a function of the detuning $\Delta/K$ for $\gamma_2=7 \gamma_1$, $E=2.25 \gamma_1$ and $K=50 \gamma_1$. We find that the position of the resonances is very well described by Eq.~\eqref{Lorentzref} (red bold dotted line). In contrast, the semi-classical model defined by Eq.~\eqref{determ} would lead to a single, broad resonance (blue dashed line). Figures~\ref{AnalytFigure} (b)-(d) show how phase locking at the two maxima and the one minimum manifests in the steady-state Wigner distribution $W(\alpha,\alpha^*)$.

Figure~\ref{Overview} (a) illustrates how more resonances at $\Delta=K(2m+1)$ appear with decreasing $\gamma_2/\gamma_1$, as more Fock levels become populated, while each individual resonance becomes weaker. The semiclassical approximation depicted in Fig.~\ref{Overview} (b) shows broadening, but there is one smeared-out resonance, as the energy distribution is continuous classically. Figure~\ref{Overview} (c) shows the synchronization tongue, i.e.~the synchronization measure as a function of detuning $\Delta$ for increasing drive $E$. The ratio $\gamma_2/\gamma_1$ is chosen such that three Fock levels have a non-negligible population in steady state resulting in the two resonances for the full quantum description. As expected classically, the tongue is not split in Fig.~\ref{Overview} (d) showing the solution for $\Lambda_c$. Finally, Figures~\ref{Overview} (e) and (f) illustrate that in the absence of a Kerr anharmonicity, $K=0$, there is only one resonance as all energy gaps are identical for harmonic oscillators. For increasing $K$ the resonance splits in the quantum system Fig.~\ref{Overview} (e), while the classical resonance Fig.~\ref{Overview} (f) broadens.

\begin{figure}[tb]
\centering
\includegraphics[width=0.5\textwidth]{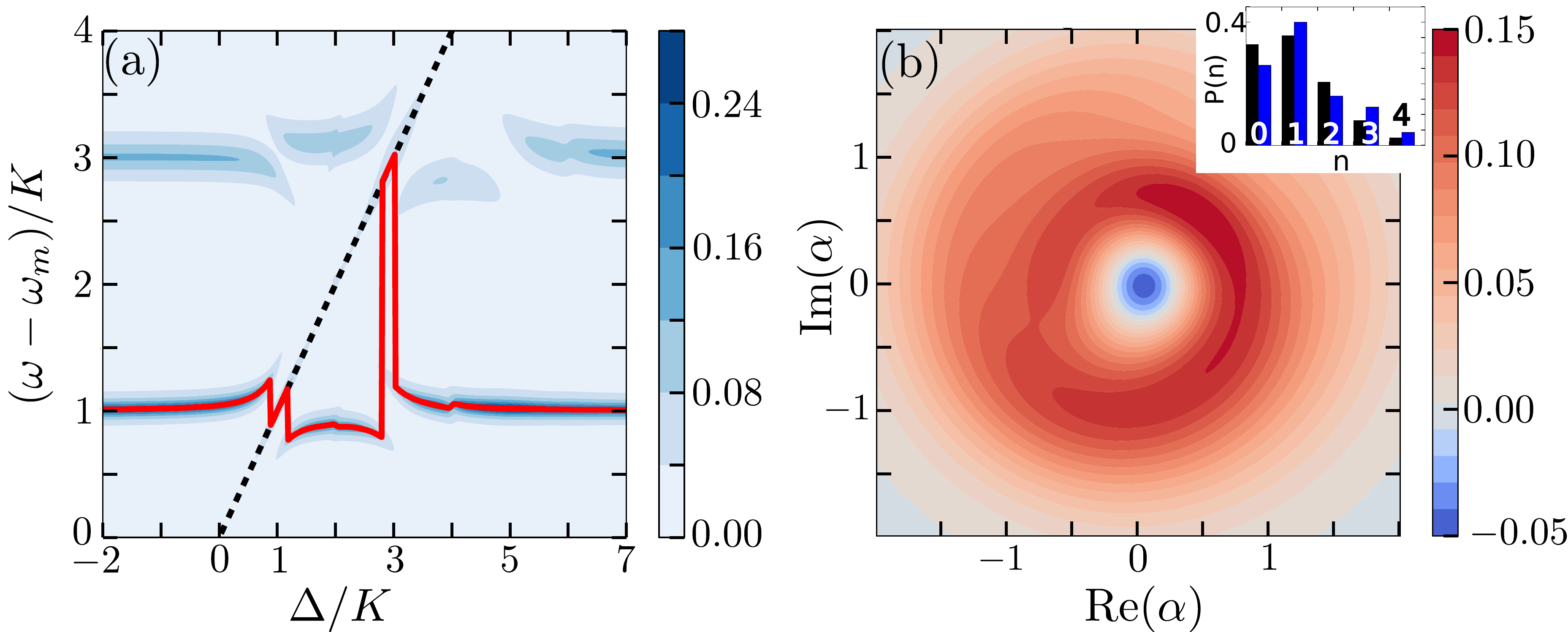}
\caption{(a) Power spectrum $P(\omega)$ of the quantum van der Pol-Kerr oscillator as a function of detuning $\Delta$ for $\gamma_2= 0.8 \gamma_1$, $E= 4.5\gamma_1$, and $K=25 \gamma_1$. The red solid line indicates the maximum of the power spectrum, the black dashed line the detuning $\Delta$ of the external drive. Around $\Delta \approx K$ and $\Delta \approx 3K$ the two lines match indicating a resonance in frequency entrainment. At $\Delta=5K$ the spectrum shows a third, smaller response. The steady state for these parameters is characterized by a Wigner distribution with negative density, see panel (b). The inset shows the Fock state probabilities $P(n)$ in the presence (right blue bars) and absence of the coherent drive (left black bars).}
\label{PowerS}
\end{figure}

\textit{Frequency entrainment and negative Wigner density}.-- We now use the power spectrum 
\begin{equation}
P(\omega)=\int_{-\infty}^\infty   e^{i \omega t} \langle b^\dagger (t) b(0) \rangle \mathrm dt
\end{equation}
to discuss the frequency entrainment \cite{Walter2014}. 
In Fig.~\ref{PowerS} (a) we demonstrate that for a nonzero Kerr anharmonicity $K \not=0$ the frequency entrainment shows resonances at detunings $\Delta=(2n+1)K$, similar to the resonances in phase locking discussed in the previous paragraph. For the parameters of Fig.~\ref{PowerS} the drive is so strong that the dynamics goes beyond first-order perturbation theory and also diagonal matrix elements of the density matrix in steady state are changed. As shown in the inset of Fig.~\ref{PowerS} (b), for the detuning at the $\Delta=5K$ resonance the redistribution is from even to odd Fock states, which have negative Wigner density around the origin $\alpha=0$. Accordingly, the steady-state Wigner distribution shows strong negative density as shown in Fig.~\ref{PowerS} (b) and therefore cannot be described in a semiclassical picture.

This clearly demonstrates that (quantum-induced) diffusion is insufficient to describe the synchronization dynamics of anharmonic oscillators, since derivatives of higher than second order are required to bring about a negative Wigner density \cite{Risken1984} in the phase space formulation of quantum optics. Here, the higher-order derivatives stem from both the Kerr and the van der Pol nonlinearity, see Eq.~\eqref{qPDG}. Interestingly though, in the case of \textit{linear} instead of nonlinear damping, the steady-state Wigner distribution can be calculated analytically \cite{Drummond1980,Kheruntsyan1999} and it is always positive, even for $K\neq 0$. Similarly, for van der Pol oscillators without Kerr term, only positive-valued Wigner densities have been found~\cite{Lee2013,Walter2014}. These observations suggest that for harmonic driving only the \emph{combination} of a Kerr anharmonicity and a van der Pol nonlinearity results in a nonclassical steady state.

\textit{Experimental implementation}.-- 
The driven van der Pol oscillator can be implemented with trapped ions, where one-phonon gain and two-phonon loss can be engineered by driving different sidebands \cite{Lee2013}. Also our additional requirement $K \gg \Gamma_m$ is feasible, as trapping potentials with very large anharmonicities in position can be realized ~\cite{Zhao2008, Wang2011,Home2011} with almost lossless resonators, e.g.~$K=20$ kHz in Ref.~\cite{Wang2011} and typical heating rates on the order of 100 Hz \cite{Epstein2007}. For optimized systems \cite{Chiaverini2014, Goodwin2016} heating rates on the order of Hz have been reported.
We further discuss effects of a finite heating rate in the Supplemental Material \cite{Supplemental}. 
The Duffing anharmonicity $\chi(a+a^\dagger)^4$ found in trapped-ion systems may be approximated by a Kerr term for $\omega_m \gg \chi \langle a^\dagger a \rangle$ using the rotating-wave approximation (RWA). The corresponding Kerr parameter is then given by $K=6\chi$.
We numerically confirm that the RWA has a large regime of validity \cite{Supplemental}.

Synchronization can also be studied in optomechanical systems \cite{Zhang2012, Bagheri2013, Matheny2014, Shlomi2015, Zhang2015}, where as a future perspective strong Kerr anharmonicities may be engineered according to proposals \cite{Jacobs2009, Rips2012, Lu2015, Zhang2015a}.

\textit{Conclusion}.-- 
We have identified parameter regimes where synchronization of a quantum anharmonic oscillator is \emph{qualitatively} different from that in the corresponding semiclassical model. We have shown that phase locking is resonantly enhanced and suppressed due to the quantization of possible oscillation frequencies, as reflected in the extrema of the synchronization tongue of Fig.~\ref{Overview}. This behavior can be understood with a simple analytical model leading to Eq.~\eqref{Lorentzref}. Frequency entrainment can switch from unlocked to nearly locked behavior at the same resonances as shown in Fig.~\ref{PowerS} (a). A further clear signature of nonclassical dynamics is the negative density in the steady-state Wigner distribution displayed in Fig.~\ref{PowerS} (b), which is in contrast to similar systems \cite{Kheruntsyan1999}. Possible experimental realizations include trapped ion setups or other platforms with strongly anharmonic spectrum. We expect that the genuine quantum signatures discussed here will be relevant in studies of synchronization in anharmonic oscillator networks or anharmonic oscillators coupled to other quantum systems such as qubits.
 
We would like to acknowledge helpful discussions with G.~Hegi, A.~Mokhberi, R.~P.~Tiwari, S.~Walter, and S.~Willitsch. 
This work was financially supported by the Swiss SNF and the NCCR Quantum Science and Technology. A.N. holds a University Research Fellowship from the Royal Society and acknowledges support from the Winton Programme for the Physics of Sustainability.

\bibliography{vdPolKerrSynchronize}

\begin{thebibliography}{57}%
\makeatletter
\providecommand \@ifxundefined [1]{%
 \@ifx{#1\undefined}
}%
\providecommand \@ifnum [1]{%
 \ifnum #1\expandafter \@firstoftwo
 \else \expandafter \@secondoftwo
 \fi
}%
\providecommand \@ifx [1]{%
 \ifx #1\expandafter \@firstoftwo
 \else \expandafter \@secondoftwo
 \fi
}%
\providecommand \natexlab [1]{#1}%
\providecommand \enquote  [1]{``#1''}%
\providecommand \bibnamefont  [1]{#1}%
\providecommand \bibfnamefont [1]{#1}%
\providecommand \citenamefont [1]{#1}%
\providecommand \href@noop [0]{\@secondoftwo}%
\providecommand \href [0]{\begingroup \@sanitize@url \@href}%
\providecommand \@href[1]{\@@startlink{#1}\@@href}%
\providecommand \@@href[1]{\endgroup#1\@@endlink}%
\providecommand \@sanitize@url [0]{\catcode `\\12\catcode `\$12\catcode
  `\&12\catcode `\#12\catcode `\^12\catcode `\_12\catcode `\%12\relax}%
\providecommand \@@startlink[1]{}%
\providecommand \@@endlink[0]{}%
\providecommand \url  [0]{\begingroup\@sanitize@url \@url }%
\providecommand \@url [1]{\endgroup\@href {#1}{\urlprefix }}%
\providecommand \urlprefix  [0]{URL }%
\providecommand \Eprint [0]{\href }%
\providecommand \doibase [0]{http://dx.doi.org/}%
\providecommand \selectlanguage [0]{\@gobble}%
\providecommand \bibinfo  [0]{\@secondoftwo}%
\providecommand \bibfield  [0]{\@secondoftwo}%
\providecommand \translation [1]{[#1]}%
\providecommand \BibitemOpen [0]{}%
\providecommand \bibitemStop [0]{}%
\providecommand \bibitemNoStop [0]{.\EOS\space}%
\providecommand \EOS [0]{\spacefactor3000\relax}%
\providecommand \BibitemShut  [1]{\csname bibitem#1\endcsname}%
\let\auto@bib@innerbib\@empty
\bibitem [{\citenamefont {Balanov}\ \emph {et~al.}(2008)\citenamefont
  {Balanov}, \citenamefont {Janson}, \citenamefont {Postnov},\ and\
  \citenamefont {Sosnovtseva}}]{Balanov2008}%
  \BibitemOpen
  \bibfield  {author} {\bibinfo {author} {\bibfnamefont {A.}~\bibnamefont
  {Balanov}}, \bibinfo {author} {\bibfnamefont {N.}~\bibnamefont {Janson}},
  \bibinfo {author} {\bibfnamefont {D.}~\bibnamefont {Postnov}}, \ and\
  \bibinfo {author} {\bibfnamefont {O.}~\bibnamefont {Sosnovtseva}},\ }\href
  {\doibase 10.1007/978-3-540-72128-4} {\emph {\bibinfo {title}
  {{Synchronization: From Simple to Complex}}}}\ (\bibinfo  {publisher}
  {Springer},\ \bibinfo {year} {2008})\BibitemShut {NoStop}%
\bibitem [{\citenamefont {Pikovsky}\ and\ \citenamefont
  {Rosenblum}(2015)}]{Pikovsky2015}%
  \BibitemOpen
  \bibfield  {author} {\bibinfo {author} {\bibfnamefont {A.}~\bibnamefont
  {Pikovsky}}\ and\ \bibinfo {author} {\bibfnamefont {M.}~\bibnamefont
  {Rosenblum}},\ }\href
  {http://scitation.aip.org/content/aip/journal/chaos/25/9/10.1063/1.4922971}
  {\bibfield  {journal} {\bibinfo  {journal} {Chaos}\ }\textbf {\bibinfo
  {volume} {25}},\ \bibinfo {pages} {097616} (\bibinfo {year}
  {2015})}\BibitemShut {NoStop}%
\bibitem [{\citenamefont {Varela}\ \emph {et~al.}(2001)\citenamefont {Varela},
  \citenamefont {Lachaux}, \citenamefont {Rodriguez},\ and\ \citenamefont
  {Martinerie}}]{Varela2001}%
  \BibitemOpen
  \bibfield  {author} {\bibinfo {author} {\bibfnamefont {F.}~\bibnamefont
  {Varela}}, \bibinfo {author} {\bibfnamefont {J.~P.}\ \bibnamefont {Lachaux}},
  \bibinfo {author} {\bibfnamefont {E.}~\bibnamefont {Rodriguez}}, \ and\
  \bibinfo {author} {\bibfnamefont {J.}~\bibnamefont {Martinerie}},\ }\href
  {http://www.nature.com/nrn/journal/v2/n4/full/nrn0401{\_}229a.html{\#}a1}
  {\bibfield  {journal} {\bibinfo  {journal} {Nat. Rev. Neurosci.}\ }\textbf
  {\bibinfo {volume} {2}},\ \bibinfo {pages} {229} (\bibinfo {year}
  {2001})}\BibitemShut {NoStop}%
\bibitem [{\citenamefont {Fell}\ and\ \citenamefont
  {Axmacher}(2011)}]{Fell2011}%
  \BibitemOpen
  \bibfield  {author} {\bibinfo {author} {\bibfnamefont {J.}~\bibnamefont
  {Fell}}\ and\ \bibinfo {author} {\bibfnamefont {N.}~\bibnamefont
  {Axmacher}},\ }\href
  {http://www.nature.com/nrn/journal/v12/n2/full/nrn2979.html{\#}a1} {\bibfield
   {journal} {\bibinfo  {journal} {Nat. Rev. Neurosci.}\ }\textbf {\bibinfo
  {volume} {12}},\ \bibinfo {pages} {105} (\bibinfo {year} {2011})}\BibitemShut
  {NoStop}%
\bibitem [{\citenamefont {Motter}\ \emph {et~al.}(2013)\citenamefont {Motter},
  \citenamefont {Myers}, \citenamefont {Anghel},\ and\ \citenamefont
  {Nishikawa}}]{Motter2013}%
  \BibitemOpen
  \bibfield  {author} {\bibinfo {author} {\bibfnamefont {A.~E.}\ \bibnamefont
  {Motter}}, \bibinfo {author} {\bibfnamefont {S.~A.}\ \bibnamefont {Myers}},
  \bibinfo {author} {\bibfnamefont {M.}~\bibnamefont {Anghel}}, \ and\ \bibinfo
  {author} {\bibfnamefont {T.}~\bibnamefont {Nishikawa}},\ }\href
  {http://dx.doi.org/10.1038/nphys2535} {\bibfield  {journal} {\bibinfo
  {journal} {Nat. Phys.}\ }\textbf {\bibinfo {volume} {9}},\ \bibinfo {pages}
  {191} (\bibinfo {year} {2013})}\BibitemShut {NoStop}%
\bibitem [{\citenamefont {Antonio}\ \emph {et~al.}(2012)\citenamefont
  {Antonio}, \citenamefont {Zanette},\ and\ \citenamefont
  {L{\'{o}}pez}}]{Antonio2012}%
  \BibitemOpen
  \bibfield  {author} {\bibinfo {author} {\bibfnamefont {D.}~\bibnamefont
  {Antonio}}, \bibinfo {author} {\bibfnamefont {D.~H.}\ \bibnamefont
  {Zanette}}, \ and\ \bibinfo {author} {\bibfnamefont {D.}~\bibnamefont
  {L{\'{o}}pez}},\ }\href {\doibase 10.1038/ncomms1813} {\bibfield  {journal}
  {\bibinfo  {journal} {Nat. Commun.}\ }\textbf {\bibinfo {volume} {3}},\
  \bibinfo {pages} {806} (\bibinfo {year} {2012})}\BibitemShut {NoStop}%
\bibitem [{\citenamefont {Arroyo}\ and\ \citenamefont
  {Zanette}(2013)}]{Arroyo2013}%
  \BibitemOpen
  \bibfield  {author} {\bibinfo {author} {\bibfnamefont {S.~I.}\ \bibnamefont
  {Arroyo}}\ and\ \bibinfo {author} {\bibfnamefont {D.~H.}\ \bibnamefont
  {Zanette}},\ }\href {\doibase 10.1103/PhysRevE.87.052910} {\bibfield
  {journal} {\bibinfo  {journal} {Phys. Rev. E}\ }\textbf {\bibinfo {volume}
  {87}},\ \bibinfo {pages} {052910} (\bibinfo {year} {2013})}\BibitemShut
  {NoStop}%
\bibitem [{\citenamefont {Dykman}(2012)}]{Dykman2012}%
  \BibitemOpen
  \bibfield  {author} {\bibinfo {author} {\bibfnamefont {M.}~\bibnamefont
  {Dykman}},\ }\href {https://books.google.com/books?id=tj0uVYnXLcUC{\&}pgis=1}
  {\emph {\bibinfo {title} {{Fluctuating Nonlinear Oscillators: From
  Nanomechanics to Quantum Superconducting Circuits}}}}\ (\bibinfo  {publisher}
  {OUP Oxford},\ \bibinfo {year} {2012})\BibitemShut {NoStop}%
\bibitem [{\citenamefont {Safavi-Naeini}\ \emph {et~al.}(2014)\citenamefont
  {Safavi-Naeini}, \citenamefont {Hill}, \citenamefont {Meenehan},
  \citenamefont {Chan}, \citenamefont {Gr{\"{o}}blacher},\ and\ \citenamefont
  {Painter}}]{Safavi-Naeini2014a}%
  \BibitemOpen
  \bibfield  {author} {\bibinfo {author} {\bibfnamefont {A.~H.}\ \bibnamefont
  {Safavi-Naeini}}, \bibinfo {author} {\bibfnamefont {J.~T.}\ \bibnamefont
  {Hill}}, \bibinfo {author} {\bibfnamefont {S.}~\bibnamefont {Meenehan}},
  \bibinfo {author} {\bibfnamefont {J.}~\bibnamefont {Chan}}, \bibinfo {author}
  {\bibfnamefont {S.}~\bibnamefont {Gr{\"{o}}blacher}}, \ and\ \bibinfo
  {author} {\bibfnamefont {O.}~\bibnamefont {Painter}},\ }\href
  {http://link.aps.org/doi/10.1103/PhysRevLett.112.153603} {\bibfield
  {journal} {\bibinfo  {journal} {Phys. Rev. Lett.}\ }\textbf {\bibinfo
  {volume} {112}},\ \bibinfo {pages} {153603} (\bibinfo {year}
  {2014})}\BibitemShut {NoStop}%
\bibitem [{\citenamefont {Zhirov}\ and\ \citenamefont
  {Shepelyansky}(2006)}]{Shepelyansky2006}%
  \BibitemOpen
  \bibfield  {author} {\bibinfo {author} {\bibfnamefont {O.}~\bibnamefont
  {Zhirov}}\ and\ \bibinfo {author} {\bibfnamefont {D.}~\bibnamefont
  {Shepelyansky}},\ }\href@noop {} {\bibfield  {journal} {\bibinfo  {journal}
  {Eur. Phys. J. D}\ }\textbf {\bibinfo {volume} {38}},\ \bibinfo {pages} {375}
  (\bibinfo {year} {2006})}\BibitemShut {NoStop}%
\bibitem [{\citenamefont {Zhang}\ \emph {et~al.}(2012)\citenamefont {Zhang},
  \citenamefont {Wiederhecker}, \citenamefont {Manipatruni}, \citenamefont
  {Barnard}, \citenamefont {McEuen},\ and\ \citenamefont {Lipson}}]{Zhang2012}%
  \BibitemOpen
  \bibfield  {author} {\bibinfo {author} {\bibfnamefont {M.}~\bibnamefont
  {Zhang}}, \bibinfo {author} {\bibfnamefont {G.~S.}\ \bibnamefont
  {Wiederhecker}}, \bibinfo {author} {\bibfnamefont {S.}~\bibnamefont
  {Manipatruni}}, \bibinfo {author} {\bibfnamefont {A.}~\bibnamefont
  {Barnard}}, \bibinfo {author} {\bibfnamefont {P.}~\bibnamefont {McEuen}}, \
  and\ \bibinfo {author} {\bibfnamefont {M.}~\bibnamefont {Lipson}},\ }\href
  {\doibase 10.1103/PhysRevLett.109.233906} {\bibfield  {journal} {\bibinfo
  {journal} {Phys. Rev. Lett.}\ }\textbf {\bibinfo {volume} {109}},\ \bibinfo
  {pages} {233906} (\bibinfo {year} {2012})}\BibitemShut {NoStop}%
\bibitem [{\citenamefont {Bagheri}\ \emph {et~al.}(2013)\citenamefont
  {Bagheri}, \citenamefont {Poot}, \citenamefont {Fan}, \citenamefont
  {Marquardt},\ and\ \citenamefont {Tang}}]{Bagheri2013}%
  \BibitemOpen
  \bibfield  {author} {\bibinfo {author} {\bibfnamefont {M.}~\bibnamefont
  {Bagheri}}, \bibinfo {author} {\bibfnamefont {M.}~\bibnamefont {Poot}},
  \bibinfo {author} {\bibfnamefont {L.}~\bibnamefont {Fan}}, \bibinfo {author}
  {\bibfnamefont {F.}~\bibnamefont {Marquardt}}, \ and\ \bibinfo {author}
  {\bibfnamefont {H.~X.}\ \bibnamefont {Tang}},\ }\href {\doibase
  10.1103/PhysRevLett.111.213902} {\bibfield  {journal} {\bibinfo  {journal}
  {Phys. Rev. Lett.}\ }\textbf {\bibinfo {volume} {111}},\ \bibinfo {pages}
  {213902} (\bibinfo {year} {2013})}\BibitemShut {NoStop}%
\bibitem [{\citenamefont {Matheny}\ \emph {et~al.}(2014)\citenamefont
  {Matheny}, \citenamefont {Grau}, \citenamefont {Villanueva}, \citenamefont
  {Karabalin}, \citenamefont {Cross},\ and\ \citenamefont
  {Roukes}}]{Matheny2014}%
  \BibitemOpen
  \bibfield  {author} {\bibinfo {author} {\bibfnamefont {M.~H.}\ \bibnamefont
  {Matheny}}, \bibinfo {author} {\bibfnamefont {M.}~\bibnamefont {Grau}},
  \bibinfo {author} {\bibfnamefont {L.~G.}\ \bibnamefont {Villanueva}},
  \bibinfo {author} {\bibfnamefont {R.~B.}\ \bibnamefont {Karabalin}}, \bibinfo
  {author} {\bibfnamefont {M.~C.}\ \bibnamefont {Cross}}, \ and\ \bibinfo
  {author} {\bibfnamefont {M.~L.}\ \bibnamefont {Roukes}},\ }\href {\doibase
  10.1103/PhysRevLett.112.014101} {\bibfield  {journal} {\bibinfo  {journal}
  {Phys. Rev. Lett.}\ }\textbf {\bibinfo {volume} {112}},\ \bibinfo {pages}
  {014101} (\bibinfo {year} {2014})}\BibitemShut {NoStop}%
\bibitem [{\citenamefont {Shlomi}\ \emph {et~al.}(2015)\citenamefont {Shlomi},
  \citenamefont {Yuvaraj}, \citenamefont {Baskin}, \citenamefont {Suchoi},
  \citenamefont {Winik},\ and\ \citenamefont {Buks}}]{Shlomi2015}%
  \BibitemOpen
  \bibfield  {author} {\bibinfo {author} {\bibfnamefont {K.}~\bibnamefont
  {Shlomi}}, \bibinfo {author} {\bibfnamefont {D.}~\bibnamefont {Yuvaraj}},
  \bibinfo {author} {\bibfnamefont {I.}~\bibnamefont {Baskin}}, \bibinfo
  {author} {\bibfnamefont {O.}~\bibnamefont {Suchoi}}, \bibinfo {author}
  {\bibfnamefont {R.}~\bibnamefont {Winik}}, \ and\ \bibinfo {author}
  {\bibfnamefont {E.}~\bibnamefont {Buks}},\ }\href {\doibase
  10.1103/PhysRevE.91.032910} {\bibfield  {journal} {\bibinfo  {journal} {Phys.
  Rev. E}\ }\textbf {\bibinfo {volume} {91}},\ \bibinfo {pages} {032910}
  (\bibinfo {year} {2015})}\BibitemShut {NoStop}%
\bibitem [{\citenamefont {Zhang}\ \emph
  {et~al.}(2015{\natexlab{a}})\citenamefont {Zhang}, \citenamefont {Shah},
  \citenamefont {Cardenas},\ and\ \citenamefont {Lipson}}]{Zhang2015}%
  \BibitemOpen
  \bibfield  {author} {\bibinfo {author} {\bibfnamefont {M.}~\bibnamefont
  {Zhang}}, \bibinfo {author} {\bibfnamefont {S.}~\bibnamefont {Shah}},
  \bibinfo {author} {\bibfnamefont {J.}~\bibnamefont {Cardenas}}, \ and\
  \bibinfo {author} {\bibfnamefont {M.}~\bibnamefont {Lipson}},\ }\href
  {http://journals.aps.org/prl/abstract/10.1103/PhysRevLett.115.163902}
  {\bibfield  {journal} {\bibinfo  {journal} {Phys. Rev. Lett.}\ }\textbf
  {\bibinfo {volume} {115}},\ \bibinfo {pages} {163902} (\bibinfo {year}
  {2015}{\natexlab{a}})}\BibitemShut {NoStop}%
\bibitem [{\citenamefont {Xu}\ \emph {et~al.}(2014)\citenamefont {Xu},
  \citenamefont {Tieri}, \citenamefont {Fine}, \citenamefont {Thompson},\ and\
  \citenamefont {Holland}}]{Xu2014b}%
  \BibitemOpen
  \bibfield  {author} {\bibinfo {author} {\bibfnamefont {M.}~\bibnamefont
  {Xu}}, \bibinfo {author} {\bibfnamefont {D.~A.}\ \bibnamefont {Tieri}},
  \bibinfo {author} {\bibfnamefont {E.~C.}\ \bibnamefont {Fine}}, \bibinfo
  {author} {\bibfnamefont {J.~K.}\ \bibnamefont {Thompson}}, \ and\ \bibinfo
  {author} {\bibfnamefont {M.~J.}\ \bibnamefont {Holland}},\ }\href
  {http://journals.aps.org/prl/abstract/10.1103/PhysRevLett.113.154101}
  {\bibfield  {journal} {\bibinfo  {journal} {Phys. Rev. Lett.}\ }\textbf
  {\bibinfo {volume} {113}},\ \bibinfo {pages} {154101} (\bibinfo {year}
  {2014})}\BibitemShut {NoStop}%
\bibitem [{\citenamefont {Xu}\ and\ \citenamefont {Holland}(2015)}]{Xu2015}%
  \BibitemOpen
  \bibfield  {author} {\bibinfo {author} {\bibfnamefont {M.}~\bibnamefont
  {Xu}}\ and\ \bibinfo {author} {\bibfnamefont {M.~J.}\ \bibnamefont
  {Holland}},\ }\href
  {http://journals.aps.org/prl/abstract/10.1103/PhysRevLett.114.103601}
  {\bibfield  {journal} {\bibinfo  {journal} {Phys. Rev. Lett.}\ }\textbf
  {\bibinfo {volume} {114}},\ \bibinfo {pages} {103601} (\bibinfo {year}
  {2015})}\BibitemShut {NoStop}%
\bibitem [{\citenamefont {Zhu}\ \emph {et~al.}(2015{\natexlab{a}})\citenamefont
  {Zhu}, \citenamefont {Schachenmayer}, \citenamefont {Xu}, \citenamefont
  {Herrera}, \citenamefont {Restrepo}, \citenamefont {Holland},\ and\
  \citenamefont {Rey}}]{Xu2015c}%
  \BibitemOpen
  \bibfield  {author} {\bibinfo {author} {\bibfnamefont {B.}~\bibnamefont
  {Zhu}}, \bibinfo {author} {\bibfnamefont {J.}~\bibnamefont {Schachenmayer}},
  \bibinfo {author} {\bibfnamefont {M.}~\bibnamefont {Xu}}, \bibinfo {author}
  {\bibfnamefont {F.}~\bibnamefont {Herrera}}, \bibinfo {author} {\bibfnamefont
  {J.~G.}\ \bibnamefont {Restrepo}}, \bibinfo {author} {\bibfnamefont {M.~J.}\
  \bibnamefont {Holland}}, \ and\ \bibinfo {author} {\bibfnamefont {A.~M.}\
  \bibnamefont {Rey}},\ }\href
  {http://stacks.iop.org/1367-2630/17/i=8/a=083063} {\bibfield  {journal}
  {\bibinfo  {journal} {New Journal of Physics}\ }\textbf {\bibinfo {volume}
  {17}},\ \bibinfo {pages} {083063} (\bibinfo {year}
  {2015}{\natexlab{a}})}\BibitemShut {NoStop}%
\bibitem [{\citenamefont {Carmele}\ \emph {et~al.}(2013)\citenamefont
  {Carmele}, \citenamefont {Kabuss}, \citenamefont {Schulze}, \citenamefont
  {Reitzenstein},\ and\ \citenamefont {Knorr}}]{Carmele2013}%
  \BibitemOpen
  \bibfield  {author} {\bibinfo {author} {\bibfnamefont {A.}~\bibnamefont
  {Carmele}}, \bibinfo {author} {\bibfnamefont {J.}~\bibnamefont {Kabuss}},
  \bibinfo {author} {\bibfnamefont {F.}~\bibnamefont {Schulze}}, \bibinfo
  {author} {\bibfnamefont {S.}~\bibnamefont {Reitzenstein}}, \ and\ \bibinfo
  {author} {\bibfnamefont {A.}~\bibnamefont {Knorr}},\ }\href {\doibase
  10.1103/PhysRevLett.110.013601} {\bibfield  {journal} {\bibinfo  {journal}
  {Phys. Rev. Lett.}\ }\textbf {\bibinfo {volume} {110}},\ \bibinfo {pages}
  {013601} (\bibinfo {year} {2013})}\BibitemShut {NoStop}%
\bibitem [{\citenamefont {Ludwig}\ and\ \citenamefont
  {Marquardt}(2013)}]{Ludwig2013}%
  \BibitemOpen
  \bibfield  {author} {\bibinfo {author} {\bibfnamefont {M.}~\bibnamefont
  {Ludwig}}\ and\ \bibinfo {author} {\bibfnamefont {F.}~\bibnamefont
  {Marquardt}},\ }\href {\doibase 10.1103/PhysRevLett.111.073603} {\bibfield
  {journal} {\bibinfo  {journal} {Phys. Rev. Lett.}\ }\textbf {\bibinfo
  {volume} {111}},\ \bibinfo {pages} {073603} (\bibinfo {year}
  {2013})}\BibitemShut {NoStop}%
\bibitem [{\citenamefont {Walter}\ \emph {et~al.}(2014)\citenamefont {Walter},
  \citenamefont {Nunnenkamp},\ and\ \citenamefont {Bruder}}]{Walter2014a}%
  \BibitemOpen
  \bibfield  {author} {\bibinfo {author} {\bibfnamefont {S.}~\bibnamefont
  {Walter}}, \bibinfo {author} {\bibfnamefont {A.}~\bibnamefont {Nunnenkamp}},
  \ and\ \bibinfo {author} {\bibfnamefont {C.}~\bibnamefont {Bruder}},\ }\href
  {\doibase 10.1103/PhysRevLett.112.094102} {\bibfield  {journal} {\bibinfo
  {journal} {Phys. Rev. Lett.}\ }\textbf {\bibinfo {volume} {112}},\ \bibinfo
  {pages} {094102} (\bibinfo {year} {2014})}\BibitemShut {NoStop}%
\bibitem [{\citenamefont {Walter}\ \emph {et~al.}(2015)\citenamefont {Walter},
  \citenamefont {Nunnenkamp},\ and\ \citenamefont {Bruder}}]{Walter2014}%
  \BibitemOpen
  \bibfield  {author} {\bibinfo {author} {\bibfnamefont {S.}~\bibnamefont
  {Walter}}, \bibinfo {author} {\bibfnamefont {A.}~\bibnamefont {Nunnenkamp}},
  \ and\ \bibinfo {author} {\bibfnamefont {C.}~\bibnamefont {Bruder}},\ }\href
  {\doibase 10.1002/andp.201400144} {\bibfield  {journal} {\bibinfo  {journal}
  {Ann. Phys.}\ }\textbf {\bibinfo {volume} {527}},\ \bibinfo {pages} {131}
  (\bibinfo {year} {2015})}\BibitemShut {NoStop}%
\bibitem [{\citenamefont {Weiss}\ \emph {et~al.}(2016)\citenamefont {Weiss},
  \citenamefont {Kronwald},\ and\ \citenamefont {Marquardt}}]{Weiss2015}%
  \BibitemOpen
  \bibfield  {author} {\bibinfo {author} {\bibfnamefont {T.}~\bibnamefont
  {Weiss}}, \bibinfo {author} {\bibfnamefont {A.}~\bibnamefont {Kronwald}}, \
  and\ \bibinfo {author} {\bibfnamefont {F.}~\bibnamefont {Marquardt}},\ }\href
  {http://iopscience.iop.org/article/10.1088/1367-2630/18/1/013043} {\bibfield
  {journal} {\bibinfo  {journal} {New J. Phys.}\ }\textbf {\bibinfo {volume}
  {18}},\ \bibinfo {pages} {013043} (\bibinfo {year} {2016})}\BibitemShut
  {NoStop}%
\bibitem [{\citenamefont {Lee}\ and\ \citenamefont
  {Sadeghpour}(2013)}]{Lee2013}%
  \BibitemOpen
  \bibfield  {author} {\bibinfo {author} {\bibfnamefont {T.~E.}\ \bibnamefont
  {Lee}}\ and\ \bibinfo {author} {\bibfnamefont {H.~R.}\ \bibnamefont
  {Sadeghpour}},\ }\href {\doibase 10.1103/PhysRevLett.111.234101} {\bibfield
  {journal} {\bibinfo  {journal} {Phys. Rev. Lett.}\ }\textbf {\bibinfo
  {volume} {111}},\ \bibinfo {pages} {234101} (\bibinfo {year}
  {2013})}\BibitemShut {NoStop}%
\bibitem [{\citenamefont {Lee}\ \emph {et~al.}(2014)\citenamefont {Lee},
  \citenamefont {Chan},\ and\ \citenamefont {Wang}}]{Lee2014a}%
  \BibitemOpen
  \bibfield  {author} {\bibinfo {author} {\bibfnamefont {T.~E.}\ \bibnamefont
  {Lee}}, \bibinfo {author} {\bibfnamefont {C.-K.}\ \bibnamefont {Chan}}, \
  and\ \bibinfo {author} {\bibfnamefont {S.}~\bibnamefont {Wang}},\ }\href
  {http://www.mendeley.com/catalog/entanglement-tongue-quantum-synchronization-disordered-oscillators/}
  {\bibfield  {journal} {\bibinfo  {journal} {Phys. Rev. E}\ }\textbf {\bibinfo
  {volume} {89}},\ \bibinfo {pages} {022913} (\bibinfo {year}
  {2014})}\BibitemShut {NoStop}%
\bibitem [{\citenamefont {Hush}\ \emph {et~al.}(2015)\citenamefont {Hush},
  \citenamefont {Li}, \citenamefont {Genway}, \citenamefont {Lesanovsky},\ and\
  \citenamefont {Armour}}]{Hush2015}%
  \BibitemOpen
  \bibfield  {author} {\bibinfo {author} {\bibfnamefont {M.~R.}\ \bibnamefont
  {Hush}}, \bibinfo {author} {\bibfnamefont {W.}~\bibnamefont {Li}}, \bibinfo
  {author} {\bibfnamefont {S.}~\bibnamefont {Genway}}, \bibinfo {author}
  {\bibfnamefont {I.}~\bibnamefont {Lesanovsky}}, \ and\ \bibinfo {author}
  {\bibfnamefont {A.~D.}\ \bibnamefont {Armour}},\ }\href {\doibase
  10.1103/PhysRevA.91.061401} {\bibfield  {journal} {\bibinfo  {journal} {Phys.
  Rev. A}\ }\textbf {\bibinfo {volume} {91}},\ \bibinfo {pages} {061401}
  (\bibinfo {year} {2015})}\BibitemShut {NoStop}%
\bibitem [{\citenamefont {Jin}\ \emph {et~al.}(2013)\citenamefont {Jin},
  \citenamefont {Rossini}, \citenamefont {Fazio}, \citenamefont {Leib},\ and\
  \citenamefont {Hartmann}}]{Jin2013}%
  \BibitemOpen
  \bibfield  {author} {\bibinfo {author} {\bibfnamefont {J.}~\bibnamefont
  {Jin}}, \bibinfo {author} {\bibfnamefont {D.}~\bibnamefont {Rossini}},
  \bibinfo {author} {\bibfnamefont {R.}~\bibnamefont {Fazio}}, \bibinfo
  {author} {\bibfnamefont {M.}~\bibnamefont {Leib}}, \ and\ \bibinfo {author}
  {\bibfnamefont {M.~J.}\ \bibnamefont {Hartmann}},\ }\href {\doibase
  10.1103/PhysRevLett.110.163605} {\bibfield  {journal} {\bibinfo  {journal}
  {Phys. Rev. Lett.}\ }\textbf {\bibinfo {volume} {110}},\ \bibinfo {pages}
  {163605} (\bibinfo {year} {2013})}\BibitemShut {NoStop}%
\bibitem [{\citenamefont {Zhu}\ \emph {et~al.}(2015{\natexlab{b}})\citenamefont
  {Zhu}, \citenamefont {Schachenmayer}, \citenamefont {Xu}, \citenamefont
  {Herrera}, \citenamefont {Restrepo}, \citenamefont {Holland},\ and\
  \citenamefont {Rey}}]{Zhu2015}%
  \BibitemOpen
  \bibfield  {author} {\bibinfo {author} {\bibfnamefont {B.}~\bibnamefont
  {Zhu}}, \bibinfo {author} {\bibfnamefont {J.}~\bibnamefont {Schachenmayer}},
  \bibinfo {author} {\bibfnamefont {M.}~\bibnamefont {Xu}}, \bibinfo {author}
  {\bibfnamefont {F.}~\bibnamefont {Herrera}}, \bibinfo {author} {\bibfnamefont
  {J.~G.}\ \bibnamefont {Restrepo}}, \bibinfo {author} {\bibfnamefont {M.~J.}\
  \bibnamefont {Holland}}, \ and\ \bibinfo {author} {\bibfnamefont {A.~M.}\
  \bibnamefont {Rey}},\ }\href
  {http://stacks.iop.org/1367-2630/17/i=8/a=083063} {\bibfield  {journal}
  {\bibinfo  {journal} {New Journal of Physics}\ }\textbf {\bibinfo {volume}
  {17}},\ \bibinfo {pages} {083063} (\bibinfo {year}
  {2015}{\natexlab{b}})}\BibitemShut {NoStop}%
\bibitem [{\citenamefont {Mari}\ \emph {et~al.}(2013)\citenamefont {Mari},
  \citenamefont {Farace}, \citenamefont {Didier}, \citenamefont {Giovannetti},\
  and\ \citenamefont {Fazio}}]{Fazio2013}%
  \BibitemOpen
  \bibfield  {author} {\bibinfo {author} {\bibfnamefont {A.}~\bibnamefont
  {Mari}}, \bibinfo {author} {\bibfnamefont {A.}~\bibnamefont {Farace}},
  \bibinfo {author} {\bibfnamefont {N.}~\bibnamefont {Didier}}, \bibinfo
  {author} {\bibfnamefont {V.}~\bibnamefont {Giovannetti}}, \ and\ \bibinfo
  {author} {\bibfnamefont {R.}~\bibnamefont {Fazio}},\ }\href {\doibase
  10.1103/PhysRevLett.111.103605} {\bibfield  {journal} {\bibinfo  {journal}
  {Phys. Rev. Lett.}\ }\textbf {\bibinfo {volume} {111}},\ \bibinfo {pages}
  {103605} (\bibinfo {year} {2013})}\BibitemShut {NoStop}%
\bibitem [{\citenamefont {Ameri}\ \emph {et~al.}(2015)\citenamefont {Ameri},
  \citenamefont {Eghbali-Arani}, \citenamefont {Mari}, \citenamefont {Farace},
  \citenamefont {Kheirandish}, \citenamefont {Giovannetti},\ and\ \citenamefont
  {Fazio}}]{Fazio2015}%
  \BibitemOpen
  \bibfield  {author} {\bibinfo {author} {\bibfnamefont {V.}~\bibnamefont
  {Ameri}}, \bibinfo {author} {\bibfnamefont {M.}~\bibnamefont
  {Eghbali-Arani}}, \bibinfo {author} {\bibfnamefont {A.}~\bibnamefont {Mari}},
  \bibinfo {author} {\bibfnamefont {A.}~\bibnamefont {Farace}}, \bibinfo
  {author} {\bibfnamefont {F.}~\bibnamefont {Kheirandish}}, \bibinfo {author}
  {\bibfnamefont {V.}~\bibnamefont {Giovannetti}}, \ and\ \bibinfo {author}
  {\bibfnamefont {R.}~\bibnamefont {Fazio}},\ }\href {\doibase
  10.1103/PhysRevA.91.012301} {\bibfield  {journal} {\bibinfo  {journal} {Phys.
  Rev. A}\ }\textbf {\bibinfo {volume} {91}},\ \bibinfo {pages} {012301}
  (\bibinfo {year} {2015})}\BibitemShut {NoStop}%
\bibitem [{\citenamefont {Gardiner}\ and\ \citenamefont
  {Zoller}(2004)}]{Gardiner2004b}%
  \BibitemOpen
  \bibfield  {author} {\bibinfo {author} {\bibfnamefont {C.}~\bibnamefont
  {Gardiner}}\ and\ \bibinfo {author} {\bibfnamefont {P.}~\bibnamefont
  {Zoller}},\ }\href
  {http://www.amazon.com/Quantum-Noise-Non-Markovian-Applications-Synergetics/dp/3540223010}
  {\emph {\bibinfo {title} {{Quantum Noise}}}}\ (\bibinfo  {publisher}
  {Springer},\ \bibinfo {year} {2004})\BibitemShut {NoStop}%
\bibitem [{\citenamefont {Carmichael}(1999)}]{Carmichael}%
  \BibitemOpen
  \bibfield  {author} {\bibinfo {author} {\bibfnamefont {H.~J.}\ \bibnamefont
  {Carmichael}},\ }\href {http://www.springer.com/us/book/9783540548829} {\emph
  {\bibinfo {title} {{Statistical Methods in Quantum Optics 1}}}}\ (\bibinfo
  {publisher} {Springer},\ \bibinfo {year} {1999})\BibitemShut {NoStop}%
\bibitem [{\citenamefont {Risken}(1984)}]{Risken1984}%
  \BibitemOpen
  \bibfield  {author} {\bibinfo {author} {\bibfnamefont {H.}~\bibnamefont
  {Risken}},\ }\href@noop {} {\emph {\bibinfo {title} {The Fokker-Planck
  Equation}}}\ (\bibinfo  {publisher} {Springer},\ \bibinfo {year}
  {1984})\BibitemShut {NoStop}%
\bibitem [{\citenamefont {Katz}\ \emph {et~al.}(2008)\citenamefont {Katz},
  \citenamefont {Lifshitz}, \citenamefont {Retzker},\ and\ \citenamefont
  {Straub}}]{Katz2008}%
  \BibitemOpen
  \bibfield  {author} {\bibinfo {author} {\bibfnamefont {I.}~\bibnamefont
  {Katz}}, \bibinfo {author} {\bibfnamefont {R.}~\bibnamefont {Lifshitz}},
  \bibinfo {author} {\bibfnamefont {A.}~\bibnamefont {Retzker}}, \ and\
  \bibinfo {author} {\bibfnamefont {R.}~\bibnamefont {Straub}},\ }\href
  {http://stacks.iop.org/1367-2630/10/i=12/a=125023} {\bibfield  {journal}
  {\bibinfo  {journal} {New Journal of Physics}\ }\textbf {\bibinfo {volume}
  {10}},\ \bibinfo {pages} {125023} (\bibinfo {year} {2008})}\BibitemShut
  {NoStop}%
\bibitem [{Sup()}]{Supplemental}%
  \BibitemOpen
  \href@noop {} {}\bibinfo {note} {See Supplemental Material}\BibitemShut
  {NoStop}%
\bibitem [{\citenamefont {{Peixoto de Faria}}(2007)}]{DeFaria2005}%
  \BibitemOpen
  \bibfield  {author} {\bibinfo {author} {\bibfnamefont {J.}~\bibnamefont
  {{Peixoto de Faria}}},\ }\href
  {http://www.mendeley.com/catalog/time-evolution-classical-quantum-mechanical-versions-diffusive-anharmonic-oscillator-example-lie-alg/}
  {\bibfield  {journal} {\bibinfo  {journal} {Eur. Phys. J. D}\ }\textbf
  {\bibinfo {volume} {42}},\ \bibinfo {pages} {153} (\bibinfo {year}
  {2007})}\BibitemShut {NoStop}%
\bibitem [{\citenamefont {Walls}\ and\ \citenamefont
  {Milburn}(2007)}]{Walls2007}%
  \BibitemOpen
  \bibfield  {author} {\bibinfo {author} {\bibfnamefont {D.~F.}\ \bibnamefont
  {Walls}}\ and\ \bibinfo {author} {\bibfnamefont {G.~J.}\ \bibnamefont
  {Milburn}},\ }\href@noop {} {\emph {\bibinfo {title} {Quantum optics}}}\
  (\bibinfo  {publisher} {Springer},\ \bibinfo {year} {2007})\BibitemShut
  {NoStop}%
\bibitem [{\citenamefont {Rodrigues}\ and\ \citenamefont
  {Armour}(2010)}]{Rodrigues2010}%
  \BibitemOpen
  \bibfield  {author} {\bibinfo {author} {\bibfnamefont {D.~A.}\ \bibnamefont
  {Rodrigues}}\ and\ \bibinfo {author} {\bibfnamefont {A.~D.}\ \bibnamefont
  {Armour}},\ }\href {\doibase 10.1103/PhysRevLett.104.053601} {\bibfield
  {journal} {\bibinfo  {journal} {Phys. Rev. Lett.}\ }\textbf {\bibinfo
  {volume} {104}},\ \bibinfo {pages} {053601} (\bibinfo {year}
  {2010})}\BibitemShut {NoStop}%
\bibitem [{\citenamefont {L{\"{o}}rch}\ \emph {et~al.}(2014)\citenamefont
  {L{\"{o}}rch}, \citenamefont {Qian}, \citenamefont {Clerk}, \citenamefont
  {Marquardt},\ and\ \citenamefont {Hammerer}}]{Lorch2014}%
  \BibitemOpen
  \bibfield  {author} {\bibinfo {author} {\bibfnamefont {N.}~\bibnamefont
  {L{\"{o}}rch}}, \bibinfo {author} {\bibfnamefont {J.}~\bibnamefont {Qian}},
  \bibinfo {author} {\bibfnamefont {A.}~\bibnamefont {Clerk}}, \bibinfo
  {author} {\bibfnamefont {F.}~\bibnamefont {Marquardt}}, \ and\ \bibinfo
  {author} {\bibfnamefont {K.}~\bibnamefont {Hammerer}},\ }\href {\doibase
  10.1103/PhysRevX.4.011015} {\bibfield  {journal} {\bibinfo  {journal} {Phys.
  Rev. X}\ }\textbf {\bibinfo {volume} {4}},\ \bibinfo {pages} {011015}
  (\bibinfo {year} {2014})}\BibitemShut {NoStop}%
\bibitem [{\citenamefont {Li}\ \emph {et~al.}(2014)\citenamefont {Li},
  \citenamefont {Petruccione},\ and\ \citenamefont {Koch}}]{Li2014}%
  \BibitemOpen
  \bibfield  {author} {\bibinfo {author} {\bibfnamefont {A.~C.~Y.}\
  \bibnamefont {Li}}, \bibinfo {author} {\bibfnamefont {F.}~\bibnamefont
  {Petruccione}}, \ and\ \bibinfo {author} {\bibfnamefont {J.}~\bibnamefont
  {Koch}},\ }\href {\doibase 10.1038/srep04887} {\bibfield  {journal} {\bibinfo
   {journal} {Sci. Rep.}\ }\textbf {\bibinfo {volume} {4}},\ \bibinfo {pages}
  {4887} (\bibinfo {year} {2014})}\BibitemShut {NoStop}%
\bibitem [{\citenamefont {Dodonov}\ and\ \citenamefont
  {Mizrahi}(1997)}]{Dodonov1997}%
  \BibitemOpen
  \bibfield  {author} {\bibinfo {author} {\bibfnamefont {V.~V.}\ \bibnamefont
  {Dodonov}}\ and\ \bibinfo {author} {\bibfnamefont {S.~S.}\ \bibnamefont
  {Mizrahi}},\ }\href {http://stacks.iop.org/0305-4470/30/i=16/a=010}
  {\bibfield  {journal} {\bibinfo  {journal} {Journal of Physics A:
  Mathematical and General}\ }\textbf {\bibinfo {volume} {30}},\ \bibinfo
  {pages} {5657} (\bibinfo {year} {1997})}\BibitemShut {NoStop}%
\bibitem [{\citenamefont {Gerry}\ and\ \citenamefont
  {Knight}(2005)}]{gerry2005}%
  \BibitemOpen
  \bibfield  {author} {\bibinfo {author} {\bibfnamefont {C.}~\bibnamefont
  {Gerry}}\ and\ \bibinfo {author} {\bibfnamefont {P.}~\bibnamefont {Knight}},\
  }\href@noop {} {\emph {\bibinfo {title} {Introductory quantum optics}}}\
  (\bibinfo  {publisher} {Cambridge University Press},\ \bibinfo {year}
  {2005})\BibitemShut {NoStop}%
\bibitem [{\citenamefont {Barak}\ and\ \citenamefont
  {Ben-Aryeh}(2005)}]{Barak2005}%
  \BibitemOpen
  \bibfield  {author} {\bibinfo {author} {\bibfnamefont {R.}~\bibnamefont
  {Barak}}\ and\ \bibinfo {author} {\bibfnamefont {Y.}~\bibnamefont
  {Ben-Aryeh}},\ }\href {\doibase 10.1088/1464-4266/7/5/001} {\bibfield
  {journal} {\bibinfo  {journal} {J. Opt. B Quantum Semiclassical Opt.}\
  }\textbf {\bibinfo {volume} {7}},\ \bibinfo {pages} {123} (\bibinfo {year}
  {2005})}\BibitemShut {NoStop}%
\bibitem [{\citenamefont {Johansson}\ \emph {et~al.}(2011)\citenamefont
  {Johansson}, \citenamefont {Nation},\ and\ \citenamefont
  {Nori}}]{Johansson2011b}%
  \BibitemOpen
  \bibfield  {author} {\bibinfo {author} {\bibfnamefont {J.~R.}\ \bibnamefont
  {Johansson}}, \bibinfo {author} {\bibfnamefont {P.~D.}\ \bibnamefont
  {Nation}}, \ and\ \bibinfo {author} {\bibfnamefont {F.}~\bibnamefont
  {Nori}},\ }\href {\doibase 10.1016/j.cpc.2012.02.021} {\bibfield  {journal}
  {\bibinfo  {journal} {Comput. Phys. Commun.}\ }\textbf {\bibinfo {volume}
  {183}},\ \bibinfo {pages} {1760} (\bibinfo {year} {2011})}\BibitemShut
  {NoStop}%
\bibitem [{\citenamefont {Johansson}\ \emph {et~al.}(2013)\citenamefont
  {Johansson}, \citenamefont {Nation},\ and\ \citenamefont
  {Nori}}]{Johansson2013}%
  \BibitemOpen
  \bibfield  {author} {\bibinfo {author} {\bibfnamefont {J.}~\bibnamefont
  {Johansson}}, \bibinfo {author} {\bibfnamefont {P.}~\bibnamefont {Nation}}, \
  and\ \bibinfo {author} {\bibfnamefont {F.}~\bibnamefont {Nori}},\ }\href
  {\doibase 10.1016/j.cpc.2012.11.019} {\bibfield  {journal} {\bibinfo
  {journal} {Comput. Phys. Commun.}\ }\textbf {\bibinfo {volume} {184}},\
  \bibinfo {pages} {1234} (\bibinfo {year} {2013})}\BibitemShut {NoStop}%
\bibitem [{\citenamefont {Drummond}\ and\ \citenamefont
  {Walls}(1980)}]{Drummond1980}%
  \BibitemOpen
  \bibfield  {author} {\bibinfo {author} {\bibfnamefont {P.~D.}\ \bibnamefont
  {Drummond}}\ and\ \bibinfo {author} {\bibfnamefont {D.~F.}\ \bibnamefont
  {Walls}},\ }\href {http://iopscience.iop.org/0305-4470/13/2/034} {\bibfield
  {journal} {\bibinfo  {journal} {J. Phys. A. Math. Gen.}\ }\textbf {\bibinfo
  {volume} {13}},\ \bibinfo {pages} {725} (\bibinfo {year} {1980})}\BibitemShut
  {NoStop}%
\bibitem [{\citenamefont {Kheruntsyan}(1999)}]{Kheruntsyan1999}%
  \BibitemOpen
  \bibfield  {author} {\bibinfo {author} {\bibfnamefont {K.~V.}\ \bibnamefont
  {Kheruntsyan}},\ }\href {http://stacks.iop.org/1464-4266/1/i=2/a=005}
  {\bibfield  {journal} {\bibinfo  {journal} {Journal of Optics B: Quantum and
  Semiclassical Optics}\ }\textbf {\bibinfo {volume} {1}},\ \bibinfo {pages}
  {225} (\bibinfo {year} {1999})}\BibitemShut {NoStop}%
\bibitem [{\citenamefont {Zhao}\ and\ \citenamefont
  {Babikov}(2008)}]{Zhao2008}%
  \BibitemOpen
  \bibfield  {author} {\bibinfo {author} {\bibfnamefont {M.}~\bibnamefont
  {Zhao}}\ and\ \bibinfo {author} {\bibfnamefont {D.}~\bibnamefont {Babikov}},\
  }\href {http://link.aps.org/doi/10.1103/PhysRevA.77.012338} {\bibfield
  {journal} {\bibinfo  {journal} {Phys. Rev. A}\ }\textbf {\bibinfo {volume}
  {77}},\ \bibinfo {pages} {012338} (\bibinfo {year} {2008})}\BibitemShut
  {NoStop}%
\bibitem [{\citenamefont {Wang}\ and\ \citenamefont
  {Babikov}(2011)}]{Wang2011}%
  \BibitemOpen
  \bibfield  {author} {\bibinfo {author} {\bibfnamefont {L.}~\bibnamefont
  {Wang}}\ and\ \bibinfo {author} {\bibfnamefont {D.}~\bibnamefont {Babikov}},\
  }\href {http://link.aps.org/doi/10.1103/PhysRevA.83.022305} {\bibfield
  {journal} {\bibinfo  {journal} {Phys. Rev. A}\ }\textbf {\bibinfo {volume}
  {83}},\ \bibinfo {pages} {022305} (\bibinfo {year} {2011})}\BibitemShut
  {NoStop}%
\bibitem [{\citenamefont {Home}\ \emph {et~al.}(2011)\citenamefont {Home},
  \citenamefont {Hanneke}, \citenamefont {Jost}, \citenamefont {Leibfried},\
  and\ \citenamefont {Wineland}}]{Home2011}%
  \BibitemOpen
  \bibfield  {author} {\bibinfo {author} {\bibfnamefont {J.}~\bibnamefont
  {Home}}, \bibinfo {author} {\bibfnamefont {D.}~\bibnamefont {Hanneke}},
  \bibinfo {author} {\bibfnamefont {J.}~\bibnamefont {Jost}}, \bibinfo {author}
  {\bibfnamefont {D.}~\bibnamefont {Leibfried}}, \ and\ \bibinfo {author}
  {\bibfnamefont {D.}~\bibnamefont {Wineland}},\ }\href@noop {} {\bibfield
  {journal} {\bibinfo  {journal} {New Journal of Physics}\ }\textbf {\bibinfo
  {volume} {13}},\ \bibinfo {pages} {073026} (\bibinfo {year}
  {2011})}\BibitemShut {NoStop}%
\bibitem [{\citenamefont {Epstein}\ \emph {et~al.}(2007)\citenamefont
  {Epstein}, \citenamefont {Seidelin}, \citenamefont {Leibfried}, \citenamefont
  {Wesenberg}, \citenamefont {Bollinger}, \citenamefont {Amini}, \citenamefont
  {Blakestad}, \citenamefont {Britton}, \citenamefont {Home}, \citenamefont
  {Itano}, \citenamefont {Jost}, \citenamefont {Knill}, \citenamefont {Langer},
  \citenamefont {Ozeri}, \citenamefont {Shiga},\ and\ \citenamefont
  {Wineland}}]{Epstein2007}%
  \BibitemOpen
  \bibfield  {author} {\bibinfo {author} {\bibfnamefont {R.~J.}\ \bibnamefont
  {Epstein}}, \bibinfo {author} {\bibfnamefont {S.}~\bibnamefont {Seidelin}},
  \bibinfo {author} {\bibfnamefont {D.}~\bibnamefont {Leibfried}}, \bibinfo
  {author} {\bibfnamefont {J.~H.}\ \bibnamefont {Wesenberg}}, \bibinfo {author}
  {\bibfnamefont {J.~J.}\ \bibnamefont {Bollinger}}, \bibinfo {author}
  {\bibfnamefont {J.~M.}\ \bibnamefont {Amini}}, \bibinfo {author}
  {\bibfnamefont {R.~B.}\ \bibnamefont {Blakestad}}, \bibinfo {author}
  {\bibfnamefont {J.}~\bibnamefont {Britton}}, \bibinfo {author} {\bibfnamefont
  {J.~P.}\ \bibnamefont {Home}}, \bibinfo {author} {\bibfnamefont {W.~M.}\
  \bibnamefont {Itano}}, \bibinfo {author} {\bibfnamefont {J.~D.}\ \bibnamefont
  {Jost}}, \bibinfo {author} {\bibfnamefont {E.}~\bibnamefont {Knill}},
  \bibinfo {author} {\bibfnamefont {C.}~\bibnamefont {Langer}}, \bibinfo
  {author} {\bibfnamefont {R.}~\bibnamefont {Ozeri}}, \bibinfo {author}
  {\bibfnamefont {N.}~\bibnamefont {Shiga}}, \ and\ \bibinfo {author}
  {\bibfnamefont {D.~J.}\ \bibnamefont {Wineland}},\ }\href {\doibase
  10.1103/PhysRevA.76.033411} {\bibfield  {journal} {\bibinfo  {journal} {Phys.
  Rev. A}\ }\textbf {\bibinfo {volume} {76}},\ \bibinfo {pages} {033411}
  (\bibinfo {year} {2007})}\BibitemShut {NoStop}%
\bibitem [{\citenamefont {Chiaverini}\ and\ \citenamefont
  {Sage}(2014)}]{Chiaverini2014}%
  \BibitemOpen
  \bibfield  {author} {\bibinfo {author} {\bibfnamefont {J.}~\bibnamefont
  {Chiaverini}}\ and\ \bibinfo {author} {\bibfnamefont {J.~M.}\ \bibnamefont
  {Sage}},\ }\href {\doibase 10.1103/PhysRevA.89.012318} {\bibfield  {journal}
  {\bibinfo  {journal} {Phys. Rev. A}\ }\textbf {\bibinfo {volume} {89}},\
  \bibinfo {pages} {012318} (\bibinfo {year} {2014})}\BibitemShut {NoStop}%
\bibitem [{\citenamefont {Goodwin}\ \emph {et~al.}(2016)\citenamefont
  {Goodwin}, \citenamefont {Stutter}, \citenamefont {Thompson},\ and\
  \citenamefont {Segal}}]{Goodwin2016}%
  \BibitemOpen
  \bibfield  {author} {\bibinfo {author} {\bibfnamefont {J.~F.}\ \bibnamefont
  {Goodwin}}, \bibinfo {author} {\bibfnamefont {G.}~\bibnamefont {Stutter}},
  \bibinfo {author} {\bibfnamefont {R.~C.}\ \bibnamefont {Thompson}}, \ and\
  \bibinfo {author} {\bibfnamefont {D.~M.}\ \bibnamefont {Segal}},\ }\href
  {http://journals.aps.org/prl/abstract/10.1103/PhysRevLett.116.143002}
  {\bibfield  {journal} {\bibinfo  {journal} {Phys. Rev. Lett.}\ }\textbf
  {\bibinfo {volume} {116}},\ \bibinfo {pages} {143002} (\bibinfo {year}
  {2016})}\BibitemShut {NoStop}%
\bibitem [{\citenamefont {Jacobs}\ and\ \citenamefont
  {Landahl}(2009)}]{Jacobs2009}%
  \BibitemOpen
  \bibfield  {author} {\bibinfo {author} {\bibfnamefont {K.}~\bibnamefont
  {Jacobs}}\ and\ \bibinfo {author} {\bibfnamefont {A.~J.}\ \bibnamefont
  {Landahl}},\ }\href {http://www.ncbi.nlm.nih.gov/pubmed/19792606} {\bibfield
  {journal} {\bibinfo  {journal} {Phys. Rev. Lett.}\ }\textbf {\bibinfo
  {volume} {103}},\ \bibinfo {pages} {067201} (\bibinfo {year}
  {2009})}\BibitemShut {NoStop}%
\bibitem [{\citenamefont {Rips}\ \emph {et~al.}(2012)\citenamefont {Rips},
  \citenamefont {Kiffner}, \citenamefont {Wilson-Rae},\ and\ \citenamefont
  {Hartmann}}]{Rips2012}%
  \BibitemOpen
  \bibfield  {author} {\bibinfo {author} {\bibfnamefont {S.}~\bibnamefont
  {Rips}}, \bibinfo {author} {\bibfnamefont {M.}~\bibnamefont {Kiffner}},
  \bibinfo {author} {\bibfnamefont {I.}~\bibnamefont {Wilson-Rae}}, \ and\
  \bibinfo {author} {\bibfnamefont {M.~J.}\ \bibnamefont {Hartmann}},\
  }\href@noop {} {\bibfield  {journal} {\bibinfo  {journal} {New J. Phys.}\
  }\textbf {\bibinfo {volume} {14}},\ \bibinfo {pages} {023042} (\bibinfo
  {year} {2012})}\BibitemShut {NoStop}%
\bibitem [{\citenamefont {L{\"{u}}}\ \emph {et~al.}(2015)\citenamefont
  {L{\"{u}}}, \citenamefont {Liao}, \citenamefont {Tian},\ and\ \citenamefont
  {Nori}}]{Lu2015}%
  \BibitemOpen
  \bibfield  {author} {\bibinfo {author} {\bibfnamefont {X.-Y.}\ \bibnamefont
  {L{\"{u}}}}, \bibinfo {author} {\bibfnamefont {J.-Q.}\ \bibnamefont {Liao}},
  \bibinfo {author} {\bibfnamefont {L.}~\bibnamefont {Tian}}, \ and\ \bibinfo
  {author} {\bibfnamefont {F.}~\bibnamefont {Nori}},\ }\href
  {http://link.aps.org/doi/10.1103/PhysRevA.91.013834} {\bibfield  {journal}
  {\bibinfo  {journal} {Phys. Rev. A}\ }\textbf {\bibinfo {volume} {91}},\
  \bibinfo {pages} {013834} (\bibinfo {year} {2015})}\BibitemShut {NoStop}%
\bibitem [{\citenamefont {Zhang}\ \emph
  {et~al.}(2015{\natexlab{b}})\citenamefont {Zhang}, \citenamefont {Peng},
  \citenamefont {{\"{O}}zdemir}, \citenamefont {Liu}, \citenamefont {Jing},
  \citenamefont {L{\"{u}}}, \citenamefont {Liu}, \citenamefont {Yang},\ and\
  \citenamefont {Nori}}]{Zhang2015a}%
  \BibitemOpen
  \bibfield  {author} {\bibinfo {author} {\bibfnamefont {J.}~\bibnamefont
  {Zhang}}, \bibinfo {author} {\bibfnamefont {B.}~\bibnamefont {Peng}},
  \bibinfo {author} {\bibfnamefont {S.~K.}\ \bibnamefont {{\"{O}}zdemir}},
  \bibinfo {author} {\bibfnamefont {Y.-X.}\ \bibnamefont {Liu}}, \bibinfo
  {author} {\bibfnamefont {H.}~\bibnamefont {Jing}}, \bibinfo {author}
  {\bibfnamefont {X.-Y.}\ \bibnamefont {L{\"{u}}}}, \bibinfo {author}
  {\bibfnamefont {Y.-l.}\ \bibnamefont {Liu}}, \bibinfo {author} {\bibfnamefont
  {L.}~\bibnamefont {Yang}}, \ and\ \bibinfo {author} {\bibfnamefont
  {F.}~\bibnamefont {Nori}},\ }\href
  {http://link.aps.org/doi/10.1103/PhysRevB.92.115407} {\bibfield  {journal}
  {\bibinfo  {journal} {Phys. Rev. B}\ }\textbf {\bibinfo {volume} {92}},\
  \bibinfo {pages} {115407} (\bibinfo {year} {2015}{\natexlab{b}})}\BibitemShut
  {NoStop}%
\end{thebibliography}%

\appendix

\newpage

\section{Supplemental Material}

\setcounter{figure}{0}
\makeatletter 
\renewcommand{\thefigure}{S\arabic{figure}}

\newcounter{defcounter}
\setcounter{defcounter}{0}

\newenvironment{myequation}{%
\addtocounter{equation}{-1}
\refstepcounter{defcounter}
\renewcommand\theequation{S\thedefcounter}
\begin{equation}}
{\end{equation}}

\begin{figure}[htb]
\includegraphics[width=0.5\textwidth]{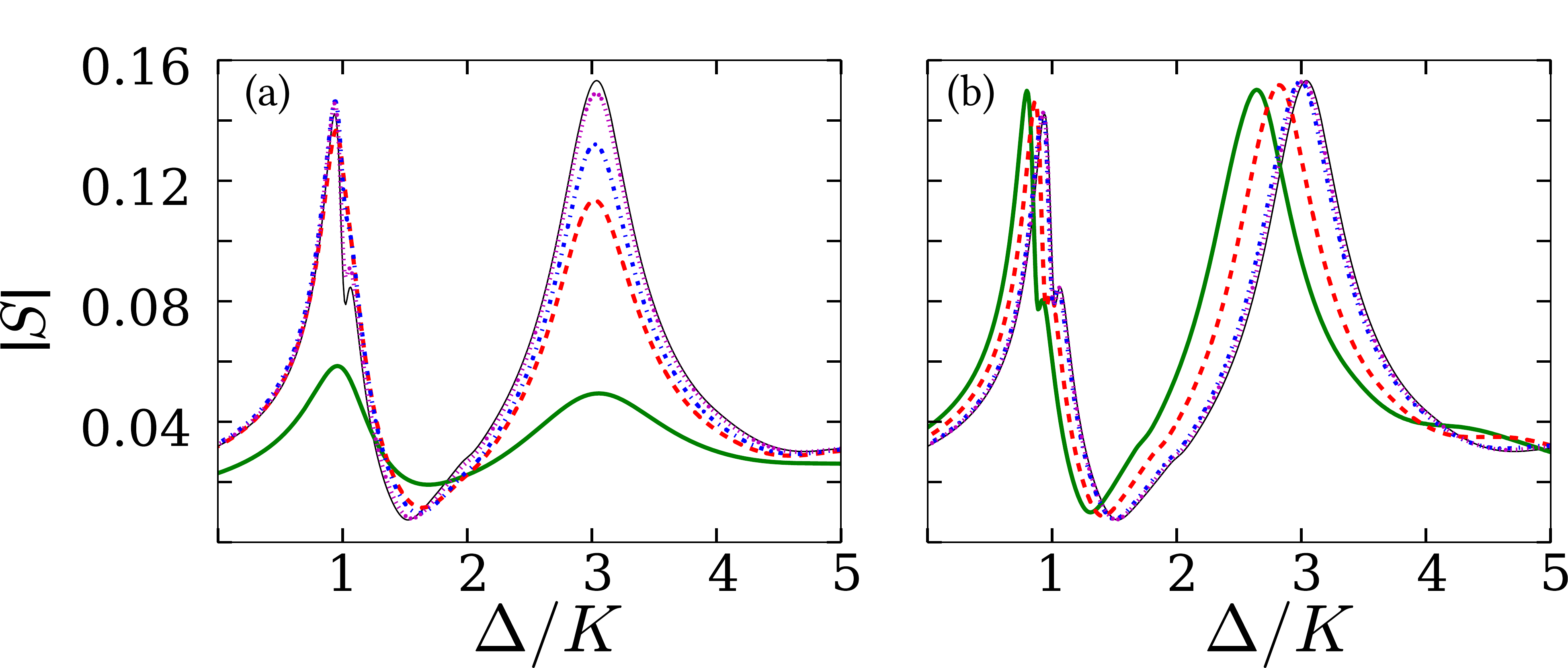}
\caption{
Synchronization measure $|S|$ in the long-time limit as a function of detuning $\Delta$ for (a) finite temperature and (b) a Duffing anharmonicity. In both plots we use the parameter set ($\gamma_2/\gamma_1=7$, $E/\gamma_1=2.25$, $K/\gamma_1=50$) from Fig. 2 of the main text and the thin black line is identical to the numerical solution of the full quantum treatment from Fig. 2, i.e. for Kerr anharmonicity and a zero-temperature bath. 
For comparison, (a) shows the numerical result for increasing strength  $\kappa \bar n$ of the heat bath in the limit $\omega_m/K \to \infty$. From top to bottom (dotted purple, dash-dotted blue, dashed red, bold green) the lines correspond to $\bar n \kappa  /\gamma_1=0.1, 0.5, 1, 5$. 
Panel (b) shows the numerical results at zero temperature for Duffing oscillators with decreasing frequency $\omega_m/K$ corresponding to larger deviations from the Kerr anharmonicity. From right to left, the lines (dotted purple, dash-dotted blue, dashed red, bold green) correspond to $\omega_m/K=1000, 500, 100, 50$. All curves where obtained by the propagator-based steady-state solver of QuTiP.
 }
\label{SuppFigure}
\end{figure}
In the main text we studied synchronization of a quantum van der Pol oscillator with Kerr anharmonicity for negligible coupling to its thermal environment. 
Here, we numerically check our predictions for the case of finite heating rates and for a Duffing type anharmonicity found e.g. in trapped ion systems as discussed in the section on experimental implementation of the main text.

We start from the quantum master equation
\begin{myequation}
\dot \rho =  -i[H_0+H_1, \rho] + (L_\gamma+L_\kappa) \rho \label{Smeq} 
\end{myequation}
in the lab frame. Therefore the drive Hamiltonian $H_1 = i E (a e^{i \omega_d t} - a^\dagger e^{-i \omega_d t})$ is time dependent, in contrast to master equation (1) from the main text, which is written in the rotating frame of the drive. 
In the Hamiltonian $H_0= \omega_m a^\dagger a  + \frac K 6 (a^\dagger+ a)^4$ the Kerr term is replaced by a Duffing anharmonicity.
The Lindblad operator $L_\gamma \rho=\frac {\gamma_1} 2 \mathcal D[a^\dagger] \rho + \frac {\gamma_2} 2 \mathcal D[a^2] \rho$  contains the van der Pol terms and corresponds to $L$ from the main text.
The new Lindblad operator $L_\kappa \rho=\bar n \frac {\kappa} 2 \left( \mathcal D[a^\dagger] \rho + \mathcal D[a] \rho \right)$ describes weak coupling to a bath of high occupation number $\bar n \gg 1$. We therfore approximate $(\bar n +1) \approx \bar n$ for the damping term. Also note that in the frame rotating at the drive frequency $\omega_d$, Eq.~(1) from the main text is obtained for $\bar n \kappa  \ll \gamma_1, \gamma_2$ and $\omega_m/K \gg 1$ by means of a rotating-wave approximation.

Since the Kerr Hamiltonian commutes with $a^\dagger a$, the master equation in the drive frame is time-independent. For the Duffing oscillator (and in general) the master equation remains time-dependent even in the rotating frame, so that also the synchronization measure $S$ remains time-dependent and enters a limit cycle at frequency $\omega_d$. To compare $S$ with the results from the main text, we could therefore either average over a period $2 \pi /\omega_d$ or compare at a particular phase of the drive in the long-time limit. 

To stay as close as possible to experiment and as the phase of the drive is generally known to the experimentalist, we choose the second option:
For times $\omega_d t= 2 \pi n$ with $n \in \mathbb N$ we show in Fig. \ref{SuppFigure} that our numerical results from the main text hold for both finite temperature (a) and finite frequency, when considering a Duffing anharmonicity (b). As expected, the resonances in the synchronization tongue may still be observed for heating rates $\bar n \kappa$ small on the scale of $\gamma_1, \gamma_2$. Similarly, for $\omega_m \gg K$, the Duffing and Kerr anharmonicity are equivalent. Even outside the regime of validity of the rotating-wave approximation, the curves are shifted to lower frequencies, but the visibility of the resonances seems unaffected.

For the ratio $K/\gamma_1=50$ used in the figure, the ion trap parameters $K=20$ kHz and $\bar n \kappa  \approx 100$ Hz \cite{Wang2011} 
correspond to $\bar n \kappa/\gamma_1=0.25$. The harmonic frequency $\omega_m=2.8$ MHz from \cite{Wang2011} fulfills $\omega_m/K>100$.
At both of these values Fig. \ref{SuppFigure} shows little deviation from the idealized setup studied in the main text. 

In conclusion, our predictions are robust to realistic heating rates and can be observed equally well if the anharmonicity stems from a Duffing potential instead of a Kerr anharmonicity.

\end{document}